\documentstyle[aas2pp4]{article}

\received{01/26/2000}

\paperid{51236}

\slugcomment{to be submitted to ApJ  ver 2000.04.26}
\newcommand{\solar}{_{\odot}} 
\newcommand{\Lxa}{L_{\rm{2-10}}^{\rm{ac}}} 
\newcommand{\Lxas}{L_{\rm{0.5-2}}^{\rm{ac}}} 
\newcommand{\Lobs}{L_{\rm{2-10}}^{\rm{obs}}} 
\newcommand{\Lint}{L_{\rm{2-10}}^{\rm{intrinsic}}} 
\newcommand{\Lints}{L_{\rm{0.5-2}}^{\rm{intrinsic}}} 
\newcommand{\Lir}{L_{\rm{FIR}}} 
\newcommand{\Lsoft}{L_{\rm{0.5-2}}^{\rm{soft}}} 
\newcommand{\Lhbeta}{L_{\rm{H\beta}}} 
\newcommand{\Lhbetascat}{L_{\rm{H\beta\_scat}}} 
\newcommand{\PBLsey}{PBL Seyfert 2 galaxies} 
\newcommand{\uflux}{ergs s$^{-1}$ cm$^{-2}$}
\newcommand{\ulum}{ergs s$^{-1}$}

\begin{document}

\title{X-ray emission from Seyfert 2 galaxies with optical polarized broad 
lines}
\author{Hisamitsu Awaki\altaffilmark{1,2}}
\affil{Department of Physics, Faculty of Science, Kyoto University}
\authoraddr{Kitashirakawa, Sakyo, Kyoto 606-8502, Japan}
\and
\author{Shiro Ueno}
\affil{Space Utilization Research Program, Tsukuba Space Center, National Space 
Development Agency of Japan,} 
\authoraddr{2-1-1 Sengen, Tsukuba 305-8505, Japan}
\and
\author{Yoshiaki Taniguchi}
\affil{Astronomical Institute, Graduate School of Science, Tohoku University}
\authoraddr{Aramaki, Aoba, Sendai 980-8578, Japan}
\and
\author{Kimberly A. Weaver}
\affil{Laboratory for High Energy Astrophysics, NASA Goddard Space Flight Center,}
\authoraddr{Greenbelt, MD, 20771, USA}

\altaffiltext{1}{CREST: Japan Science and Technology Corporation (JST), 
4-1-8 Honmachi, Kawaguchi, Saitama 332}
\altaffiltext{2}{present address: Department of Fundamental Material Science, 
Faculty of Science, Ehime University, Matsuyama, 790-8577, Japan }

\begin{abstract}

We analyze the $0.5-10$ keV spectra of six Seyfert 2 galaxies observed 
with the X-ray satellite ASCA: Mrk 3, Mrk 348, Mrk 1210, Mrk 477, NGC 7212, 
and Was 49b.  These galaxies were selected based on their possession of optical
polarized broad lines.  In the $2-10$ keV band, their spectra are heavily absorbed, 
with $2-10$ keV absorption-corrected X-ray luminosities ranging 
from 10$^{42}$ to 10$^{43}$ \ulum.  The observed X-ray emission is
generally only about one tenth that predicted based on their known infrared 
and H$\beta$ luminosities.  This apparent X-ray weakness can be 
understood if a considerable fraction of the nuclear activity is completely 
blocked from view by thick matter along our line of sight to the nucleus.

All of these galaxies possess significant soft X-ray emission whose 
origin appears to be scattered light from their nuclear emission.  Based 
on this hypothesis, we estimate
a typical scattering efficiency for X-rays to be about 10$\%$.  This
efficiency is larger than the few $\%$ found 
for ordinary Seyfert 2 galaxies with no report of optical polarized broad  
lines.  A large scattering efficiency is best explained by an apparent 
weakness of the hard X-ray luminosity rather than unusually strong scattered 
light in the soft band.  When we estimate the scattering efficiency using the
intrinsic luminosity derived assuming Compton scattering 
dominates the hard X-ray spectrum, as
opposed to a purely absorbed nuclear continuum, the efficiency can be less 
and is similar to that of ordinary Seyfert 2 galaxies.  
Since the difference between our sample and ordinary Seyfert 2 galaxies can
be explained by the difference of viewing angle,
the similar scattering efficiency suggests the existence of a scattering 
region that is larger than the putative dusty torus. 

\end{abstract}

\keywords{galaxies: active--- galaxies: Seyfert---X-rays: galaxies}

\section{INTRODUCTION}

Seyfert galaxies are nearby active galaxies that have been traditionally
classified into two types according to the presence or absence of broad 
optical/UV emission lines (more than a couple thousand km s$^{-1}$ FWHM).
However, after broad lines were discovered in the 
polarized spectrum of the typical Seyfert 2 galaxy, NGC 1068 (Antonucci
\& Miller 1985), it became clear that broad-line region
clouds can exist in Seyfert 2 galaxies as well as Seyfert 1s,
although these clouds may be located inside 
(and hidden behind) an optically and geometrically thick disk. 
With the development of better instruments and
techniques, spectropolarimetry has revealed the existence of polarized broad 
emission lines (and hence hidden BLRs) from more than ten 
Seyfert 2 galaxies (e.g. Miller \& Goodrich 1990, Tran, Miller, \& Kay 1992).  

Based on these polarized Seyfert 2 galaxies, a viewing-dependent 
unification scenario has been proposed that predicts the existence of a 
hidden active nucleus and broad line region in all Seyfert galaxies, 
making Seyfert 1 and 2 galaxies are essentially identical. 
Hard X-ray observations with the Japanese X-ray satellite 
Ginga revealed hidden luminous nuclei in several Seyfert 2 galaxies (e.g. Awaki 
et al. 1991), demonstrating that hard X-ray observations 
are useful to reveal highly obscured nuclei.  Since 
nuclear X-rays pass though the thick matter, such galaxies
are the best targets to investigate AGN environments.  In particular, it is
crucial to reveal the origin of the weak excess emission in the soft X-ray band
(e.g. Krupper, Urry \& Canizares 1990).  One possibility is that the soft 
X-rays represent scattered light, similar to optical polarized 
broad lines (Netzer, Turner \& George 1998).  
Another possibility is that soft X-rays can come from starbursts
(e.g., Dahlem, Weaver \& Heckman 1998) or 
AGN-driven jets (Colbert et al. 1998).

If X-ray and optical/UV photons are scattered in
the same way, there should be a relation between them.  
To investigate whether the soft X-ray emission in Seyfert 2 galaxies is 
related to nuclear activity, we examine those known to have optical 
polarized broad emission lines.
Choosing objects with significant polarization 
maximizes the chance of detecting scattered X-rays from the AGN 
as opposed to X-rays from the galaxy. 
The Japanese X-ray satellite ASCA allows spectroscopy in the entire 
$0.5-10$ keV band, and so we can directly measure both the hard X-ray 
emissions from obscured nuclei and any associated soft X-ray emission.
By comparing our X-ray results with those of optical spectropolarimetric 
observations, we can test the origin of the soft component and investigate 
the properties of the nuclei and scattering regions in Seyfert 2 galaxies with 
polarized broad lines  (hereafter \PBLsey).  The differences between 
the amount of polarization in \PBLsey~compared to more typical Seyfert 2 
galaxies that
have no detectable polarization, should affect their corresponding 
X-ray properties.  We attempt to clarify this difference in our analysis.

\section{TARGET SELECTION AND DATA REDUCTION}

Miller \& Goodrich (1990) and Tran, Miller \& Kay (1992) found about 
10 \PBLsey.   Eight of these have been observed with
ASCA.   ASCA data for NGC 1068, 
Mrk 463E and the first observation of Mrk 3 have been 
reported by Ueno et al (1994), Ueno et al. (1996) and Iwasawa et al. 
(1994), respectively.  We refer to these papers for these three objects,
and we analyze the other five Seyfert 2 galaxies (Mrk 348, Mrk 477, Mrk 1210, 
NGC 7212, Was 49b) and the second observation of Mrk 3. 

ASCA (Tanaka, Inoue, \& Holt 1994) has four focal-plane 
detectors: two solid-state imaging spectrometers (called here SIS 0 
and SIS 1) and two 
gas imaging spectrometers (called GIS 2 and GIS 3). 
The SIS and GIS operate simultaneously.  
In our observations, the SIS data were taken with the instrument 
in 1-CCD FAINT mode at all bit-rates.
The GIS data were obtained in PH mode at all bit-rates.  
The field of view of SIS 1-CCD mode and GIS are  11$^{\prime}$ square, 
and  40$^{\prime}$ diameter, respectively.

We select data using the following data filtering,

COR $>$ 4 GeV c$^{-1}$, ELV $>$ 5 deg, and T$\_$SAA $>$ 60 s,

\noindent
where COR is cutoff rigidity, ELV is target elevation from Earth's rim
and T$\_$SAA is the time after passage through the South Atlantic Anomaly.
In order to reject 
contaminating X-rays from Earth's rim, data are also filtered using the 
angular distance from Earth's rim during daylight (BR$\_$EARTH). 
For Mkn 3, we accept data with BR$\_$EARTH $>$ 35 deg and for the 
other targets, we accept data with BR$\_$EARTH $>$ 10 deg.
The observation log and the exposure times after filtering 
are listed in Table 1.

We accumulate SIS0 and SIS1 data within a 6$^{\prime}$ diameter circle 
centered on the X-ray source; background spectra 
are taken from a non-source region $>$ 8$^{\prime}$ in diameter.  
For the GIS, we sum data from GIS2 and GIS3 in an 8$^{\prime}$ 
diameter circle centered on the X-ray source;  
background spectra are taken from an annulus with 
12$^{\prime}$-32$^{\prime}$ 
diameter centered on the X-ray source.  
We find several X-ray sources in the background region and exclude 
contaminating sources with 8$^{\prime}$ (for SIS) and 12$^{\prime}$ (for GIS)
diameter circles. The count rates after 
background subtraction are listed in Table 1.

In the field of Was 49b, there is a bright source, RX J1214.4+2936, located
4$^{\prime}$.6 away from the target galaxy (Appenzeller et al. 
1998). Although this source is classified as a Seyfert 2 galaxy at redshift 
of 0.307, the X-ray characteristics are similar to that of a Seyfert 1 galaxy or a
QSO, i.e. the spectrum is described by a single power-law model with a photon 
index of 1.30$\pm$0.15 and an absorption column of less than 6$\times 
10^{20}$ cm$^{-2}$. The X-ray flux is $\sim$8$\times$10$^{-13}$ 
\uflux~in the $2-10$ keV band, which corresponds to an X-ray 
luminosity of $\sim$4$\times$10$^{44}$ \ulum. 
The amount of contamination from this source is estimated to be
only 10$\pm$2$\%$ of the X-ray flux of Was 49b in the $0.5-2$ keV band,
and so it should not seriously affect our analysis.

For Mrk 477, Mrk 1210, and NGC 7212, which were observed twice, we first 
analyze the two observations separately. Table 1 shows the
count rate in the $0.5-10$ keV band for each.  Since their 
count rates do not vary significantly and their hardness 
ratios of $2-10$ keV versus $0.5-2$ keV rate remain the same to 
within a 1 $\sigma$ error, we 
conclude that their X-ray spectra do not change significantly 
between the two observations.  As a result, the 
separate observations are added to produce one spectrum for
each galaxy for analysis purposes. 

We combine energy bins to contain more than 40 counts in each bin and
apply a $\chi^2$ method in the XSPEC package for spectral fitting.

\section{SPECTRAL ANALYSIS AND RESULTS}
The X-ray spectra of Seyfert 2 galaxies can be approximated by 
equation (1), which represents a two-component continuum 
plus an Fe K$\alpha$ line (e.g. Turner et al. 1997a). We 
apply this model, labeled model 1, in our fitting procedure, with a 
power-law as the soft X-ray component to represent the contribution of
scattered emission from the AGN.
Some broadening of the Fe K line is reported by Turner et al. 
(1997a), but it is unclear whether there is significant line broadening
in our data.  We thus assume that the Fe K line is narrow for all cases. 
Model 1  
describes all sources well with a reduced $\chi^2\sim$1. The fitting 
results are listed in Table 2, and the spectra are shown in Figure 1.   

\begin{eqnarray*}
I (\rm{photon~s^{-1} cm^{-2} keV^{-1}}) = \rm {exp}(-\it{N}_{\rm H0}\sigma_{\rm abs})~\times
\end{eqnarray*}
\begin{equation}
~~~(\rm{soft~component} + \rm{exp}(-\it{N}_{\rm H1}\sigma_{\rm abs})~( \it{A~E^{-\Gamma}} + \rm{line})), 
\end{equation}

\noindent
where the soft component is a power-law in models 1, 2, and 3 and 
a Raymond plasma in model 4. 

For model 1, 
the observed $2-10$ keV fluxes for the hard components range from 
3$\times$10$^{-13}$ to 1.8$\times$10$^{-12}$ \uflux, which
correspond to observed luminosities ($\Lobs$) of  $10^{42} - 
10^{43}$ \ulum~assuming $H_{\rm{0}}=50$ km s$^{-1}$ Mpc$^{-1}$ (see Table 3).

\subsection {The hard X-ray components}
Hard X-ray emission with $\it{\Gamma}\sim$2 is often seen in 
normal galaxies (e.g. Okada 1998) and is thought to originate from a 
superposition of discrete sources (e.g. low mass X-ray binaries
and SNe; Makishima et al. 1989).  It is well known that the ratio
of X-ray to blue band luminosities ($L_{\rm{B}}$) is about 
4$\times$10$^{-5}$ for normal and starburst galaxies (e.g. Awaki 1999),
i.e. the luminosity from the superposition of discrete sources is 
estimated to be 1.5$\times$10$^{39}$ ($L_{B}$/10$^{10} L\solar$) 
\ulum~in the $2-10$ keV band.
Since $L_{\rm{B}}$ for our sample is order 10$^{10} L\solar$ 
(Whittle et al. 1992), we conclude that the hard 
X-ray emission is most likely to originate from nuclear activity. 

When applying model 1, the photon indices of the hard component
are poorly determined, ranging from $-0.14$ to 2.22.  Assuming the 
hard emission is related to the AGN, we next fix the   
photon index of the hard component to be 1.7 (model 2), which
is the canonical observed value for type 1 AGNs (e.g. Mushotzky 1982).
When the best fit value of $N_{\rm{H0}}$ is less than the galactic 
absorption, we set $N_{\rm{H0}}$ equal to the galactic column.\footnote{ 
obtained from EOLS/EINLINE V 2.4 at Smithsonian Astrophysical 
Observatory.}
The equivalent 
width of the Fe K$\alpha$ line is the most important parameter 
to investigate the origin of the hard component (e.g. Awaki et al. 1991 and
Turner et al. 1997b), and so we fix the center energy of the line  
at 6.4 keV (in the rest frame), the energy of Fe K$\alpha$ fluorescence 
from cold matter.  

When fixing the hard X-ray photon indices (model 2), the values of 
$\chi^{2}$ increase because the overall spectra tend to be flatter 
than a $\it{\Gamma}\sim$1.7 spectrum.
However, if our assumptions about the intrinsic spectral shape of
the hard X-ray (AGN) component are correct, then model 2 implies 
large column densities and large 
equivalent widths for the Fe K$\alpha$ line (see Table 2). The 
absorption corrected $2-10$ keV luminosities for the hard component are
greater than 10$^{42}$ \ulum~(see Table 4; hereafter $\Lxa$ denotes
this absorption corrected luminosity).  
$\Lxa$ is useful to compare with results in the literature.

We found that some galaxies have indices from 0 to 1 in model 1, which 
are smaller than the canonical Seyfert 1 value of $\sim$1.7. Matt et al. 
(1996) point out that, besides absorption, a flat spectrum can result 
from a significant 
Compton reflection component.  To test this idea, we apply a Compton 
reflection model (model 3) to the hard 
components of Mrk 1210, Mrk 477, Was49b and NGC 7212
and find that the hard X-ray emission can be reproduced 
by Compton reflection (Table 2).  The $\chi^{2}$ values
are less than those for the absorbed power-law model (models 1 and 2),
although iron line intensities are much weaker than predicted 
for reflection by Reynolds et al. (1994) and  Matt et al. (1996).

\subsection {The Soft X-ray components}
For a power-law description of the soft  X-ray emission we find 
observed 0.5 to 2 keV fluxes of about 10$^{-13}$ \uflux~
and photon indices of $\it{\Gamma}\sim$1-2.  Absorption in  
excess of the galactic column is not indicated for
most galaxies.  The X-ray luminosities for the soft components are listed 
in Table 4. 

In order to test the origin of the soft component, we characterize 
it using the Raymond-Smith plasma model (Raymond \& Smith 1977)
(model 4). 
We set the metal abundance free, because the thermal 
emission detected from starburst galaxies and NGC 1068 can be reproduced 
by a thin thermal plasma model with sub-solar abundances (e.g. Ueno et al. 
1994 for NGC 1068, Tsuru et al. 1997 for M82).
We find the temperatures for NGC 7212 and Mrk 477 to be greater 
than several keV, while the temperature and metal abundance for Mrk 1210 and
Was 49b are 0.86 keV and 0.03, and 0.72 keV and 0.02, respectively.

It is known that some Seyfert 2 galaxies possess line-like structures
around 0.9 keV (e.g. Hayashi et al. 1996 for NGC 2110, Comastri et al. 1998 
for NGC 4507). Line emission is important to investigate the origin of 
the soft component, because highly ionized neon (Ne IX), and fully 
ionized oxygen make a line-like structure at about 0.9 keV (e.g. Netzer 
1995).  We examined the line emission around 0.9 keV adding a narrow line 
to model 2.  Line-like features are apparent in the spectra of Mrk 3 and 
Mrk 477 (Table 5). The detection of the line in the spectrum for Mrk 3
is consistent with the result by Griffiths et al. (1998).     

\section{DISCUSSION}
\subsection{The Hard Component}
\subsubsection{Nuclear Activities in Seyfert 2 Galaxies with Polarized Broad Lines}

Based on our spectral fits, we find hard X-ray emission associated with nuclear 
activities in \PBLsey.   For an absorbed power-law model with a photon index 
of 1.7 (our model 2), we infer X-ray activities with $\Lxa 
> 10^{42}$ \ulum~obscured by thick matter with column densities 
of $\sim$10$^{23}$ cm$^{-2}$.  Assembling the available X-ray results for 
other \PBLsey, (NGC 1068; Ueno et al. 1994, 
NGC 7674; Malaguti et al. 1998, and Mrk 463E; Ueno et al. 1996), 
obscured nuclei are detected in all \PBLsey~observed 
to date in the hard X-ray band.  We conclude that obscured nuclei are  
common in \PBLsey.

Using $\Lxa$, we can examine whether the estimated number of photons
are sufficient to ionize the surrounding region.
For Seyfert 1 galaxies, the X-ray luminosity in the 
$2-10$ keV band ($L_{\rm{X}}$) is well correlated with the broad $H\beta$ 
luminosity ($\Lhbeta$),
with a ratio of $L_{\rm{X}}$/$\Lhbeta\sim$ $10-100$ (e.g. 
Blumenthal, Keel, \& Miller 1982).  Figure 2 shows this correlation   
for Seyfert 1 galaxies and QSOs, quoted from 
Padovani \& Rafanelli (1988) and Malaguti, Bassam, \& Caroli (1994), 
respectively (solid circles).  If Seyfert 2 galaxies are 
Seyfert 1s viewed edge-on, their intrinsic properties should 
be the same as Seyfert 1s.  We test this idea by plotting the \PBLsey\ (open circles)
in Figure 2, estimating $\Lhbeta$ from the polarized $H\beta$ flux, $\tau$=0.1,
$\Delta\Omega/4\pi$=0.1, and known polarization degree (Miller \& Goodrich 1990, 
Tran 1995a).  It is clear that 
$\Lxa$ for \PBLsey~lies systematically below the relation 
for Seyfert 1 galaxies and QSOs, such that there are 10 times fewer
ionizing photons than expected, assuming
our estimate of $\Lhbeta$ is correct.  The discrepancy is similar
for other indicators of nuclear activity;
$\Lxa$/$L_{\rm{IR}}$ for \PBLsey~are only 1/10 that
of Seyfert 1 galaxies (Figure 3a)\footnote{The IRAS color for \PBLsey~are
warm (log ($F_{\rm{25\mu m}}$/$F_{\rm{60\mu m}} > $-0.5),
which indicates that the dominant energy source of
IR emission is AGN activity.}, and Bassani
et al. (1999) point out a small ratio of $\Lxa$ to the luminosity of [OIII].
We conclude that correcting only for hard X-ray absorption results in 
\PBLsey~having $\Lxa$ that is 10 times less than
predicted if they harbor a Seyfert 1 nucleus.

\subsubsection{Weak Activity?}
If models 1 or 2 do not represent the actual spectral 
form of the nuclear source, $\Lxa$ does not represent the true luminosity.
We found that the X-ray spectra for some PBL Seyfert 2s are well represented
with the Compton reflection models, although their iron emission lines are 
weaker than that for a pure reflection model (e.g. Matt et al. 1996).  
Since most of the iron emission in this scenario is attributed to the 
reflection component, the 
contribution of reflection to the hard component can be
estimated from a ratio of the measured to predicted iron line 
intensities.  Assuming 
the equivalent width of 1.5 keV for a pure reflection component, the 
contribution is estimated to be $20-50\%$.
From theoretical considerations, the upper limit to the efficiency of 
the Compton mirror in the 2--10 keV band is about $7\%$ for 
$\Delta\Omega/2\pi$ = 1.0.  If we assume an efficiency of 5$\%$, the 
true X-ray luminosity ($\Lint$) is described as 

\begin{equation}
 \Lint = 20\times\Lxa (\frac{EW}{1500 eV}) (\frac{Eff_{\rm{Compton}}}{0.05})^{-1},
\end{equation}

\noindent
where EW and Eff$_{\rm{Compton}}$ are the observed equivalent width of 
iron K$\alpha$ line and the efficiency of the Compton mirror in the 
2--10 keV band respectively. Figure 
3b shows a plot of $\Lint$ versus $\Lir$.  \PBLsey~have the same 
ratio of $\Lint/\Lir$ as Seyfert 1s.  Therefore, by accounting 
for the observed spectrum 
as reflected rather than absorbed emission,
the intrinsic source would be ten times more luminous and  
\PBLsey~ would be consistent with having powerful AGN activity.  
Tran 1995b pointed out the existence of strong nuclear activities
in NGC 7212 and Mrk 463E from the line ratio diagnostic of jet regions. 
The powerful activities inferred from the X-ray data after the correction 
of reflection efficiency 
are consistent with his claim.

\subsubsection{Time Variability}
Seyfert 1 galaxies with X-ray luminosities of 10$^{43}$ \ulum~have 
variability time scales of 10000 seconds.  If \PBLsey~ possess obscured
Seyfert 1 type nuclei but we are seeing their hard X-ray 
emission directly, similar time variability is expected.
However, Mrk 1210, Mrk 477 and NGC 7212 show no significant 
time variability in periods of one month, one day and one week, respectively. 
This probably indicates that these galaxies do not have small 
variability time scale such as Seyfert 1 galaxies, although we need more 
X-ray monitors in order to achieve the conclusion. 
Mrk 3 has been monitored with the Einstein, Ginga, BBXRT, ROSAT
and ASCA X-ray missions.  Figure 4 shows light curves in the
soft and hard bands.  As Marshall et al. (1991) and Iwasawa et al. (1994) 
mentioned, the hard X-ray flux gradually decreases between the Ginga and 
ASCA observations.  The variability time scale is about 1 yr, which is
similar to that of luminous QSOs.

A lack of short timescale spectral variability for \PBLsey~can be explained by 
more massive black holes with smaller accretion rates and/or the dominance
of the Compton reflection process for energies approaching 10 keV.  The latter  
indicates more powerful X-ray activity, and so both possibilities suggest
the existence of more massive black holes than those of Seyfert 1 galaxies with
the same apparent X-ray luminosity.  This result is consistent with 
arguments by Nishiura \& Taniguchi (1998), who deduced large central black hole
masses from analysis of line widths of polarized broad lines.

\subsection{The Soft X-ray Component}
\subsubsection {The Origin of the Soft component} 

We detect distinct soft X-ray components with  
X-ray luminosities in the $0.5-2$ keV band of about 10$^{41}$ \ulum.  
No associated strong absorption indicates that the 
X-rays originate outside the thick matter that absorbs the hard X-rays.
The soft X-rays can be associated with circumnuclear starburst activities 
which are often seen in Seyfert 2 galaxies (e.g. Maiolino \& Rieke 1995).
On the other hand, optical scattered light is observed by 
spectropolarimetry.  From the 
analogy of these observations, it is naturally suggested that X-rays 
from nuclei are also scattered into our line of sight. 
We examine possibilities for the origin of the soft component
using 1) the X-ray spectrum, 2) the correlation between Lx 
and $\Lir$, and 3) the correlation
between Lx and luminosities of scattered $H\beta$ lines.

The X-ray spectrum:
we tried to fit the soft components with either the absorbed power-law model 
(models 1 and 2) or the thin thermal plasma model (model 4).  
NGC 7212, Mrk 477, Mrk 3, and Mrk 348 have flat spectra represented 
by a power-law model with a photon index $<$ 2.3. 
This result, combined with little soft X-ray absorption,
suggests that thermal emission due to starburst activity is 
not dominant in the soft X-ray band.
In addition, we detect a line-like feature around 0.9 keV for Mrk 3
and Mrk 477, which is one characteristic of a scattered light in
a highly ionized plasma region (Netzer 1995, Griffiths et al. 1998).  

On the other hand,  Mrk 1210 and Was 49b have softer spectra than those
of the other targets.  Their soft components can be 
described as thin thermal emission with low metal abundances of 
$<$0.1 and $<$0.14 having $kT\sim$0.86 keV and $kT\sim$0.72 keV, 
respectively. For Mrk 463E,
Ueno et al (1994) report a thermal soft X-ray component with $kT\sim$1 keV.
Since Mrk 463E and Was 49b are galaxy-margers, there is a possibility of
on-going starburst activity. However, there are 
no reports on the existence of powerful starburst activities in their nuclear 
regions, and their infrared colors (log $F_{\rm{25\mu m}}$/$F_{\rm{60\mu m}}$) 
in the IRAS band are about 0.
Therefore, we consider that most soft X-rays are associated with 
their AGN activities.  We note that a possibility of the starburst origin 
can not be ruled out from our spectral analysis alone.

$\Lir$ vs $L_{\rm{X}}$:
the X-ray luminosities of optically thin thermal emission 
from starburst galaxies and 
normal galaxies are well correlated with far infrared luminosity. 
 David, Jones \& Forman (1992) found the strong correlation with
$\Lir$/$L_{\rm{X}}\sim$2000 in the 0.5--4.5 keV band from a detail analysis 
for a large number of galaxy samples observed by Einstein.
  Figure 5 presents our sample.
We find that most Seyfert 2 galaxies have excess soft emission, and that
the contribution of the starburst component is estimated to be less than 
50$\%$ except for NGC 1068.  We note that NGC 1068 is a luminous infrared 
source with powerful starburst activity.  Thus, the soft 
components for our target galaxies are most likely to originate from 
their AGN activities rather than starburst activities.  

$\Lhbeta$ vs $L_{\rm{X}}$:
it is known that the X-ray luminosities of Seyfert 1 
galaxies are well correlated with  $\Lhbeta$ 
(Blumenthal, Keel, \& Miller. 1982) because broad $H\beta$ lines are 
emitted from photoionized regions.   If the broad $H\beta$ lines 
and nuclear light are scattered in the same region, the luminosity of 
the scattered $H\beta$ lines should correlate with that of the 
scattered X-ray component.
Figure 6 shows a correlation of the soft component and the
scattered broad $H\beta$ line luminosities ($\Lhbetascat$) for our 
sample.  The $\Lhbetascat$ were estimated using eq.(3.1) in Tran (1995b).
The two solid lines in Figure 6 show an extrapolation from the
relation found in Seyfert 1 galaxies (see section 4.1).  Considering 
the uncertainty of polarization degree, the objects in our sample 
lie in the area of Seyfert 1 galaxies.   The consistency of properties
between our sample and Seyfert 1 galaxies indicates that most of 
the soft X-rays in \PBLsey~can be explained by scattered light, 
and that electron
scattering dominates dust scattering for our sample.  
This result is consistent with early work by Tran (1995b).  

From these relevant considerations, 
we conclude that soft X-rays in \PBLsey~originate mostly from scattered
light.  We note that Heckman et al. (1997) point out that Mrk 477 has a
powerful nuclear starburst with a bolometric luminosity of $3\times10^{10}
- 10^{11} L\solar$.  The luminosity due to the starburst activity is
comparable to that of the AGN activity, however it appears to be dominated 
by the AGN in soft X-rays (Levenson, Weaver and Heckman, in prep.). 
Using the ratio of $\Lir$/$L_{\rm{X}}$ for starburst galaxies, the 
contribution of the
starburst activity in the soft band is estimated to be about 
10$^{40}$ \ulum, which is about 10$\%$ of the observed luminosity.  
Therefore we conclude that we can ignore the 
contribution of the starburst activity in Mrk 477.

\subsubsection{Scattering Efficiency of the Warm Mirror}

Using the wide band ASCA spectra, we can estimate the scattered and 
intrinsic X-ray luminosities at the same time.  If scattered light 
dominates the soft X-ray component for \PBLsey, we can use the ratio of 
normalizations between the soft and the hard components to indicate 
the scattering efficiency in the scatterers.  To do this for the continuum  
spectra, we set $\it{\Gamma}$(soft) equal to $\it{\Gamma}$(hard). 
However, there are several spectral lines in the scattered 
component (e.g. Netzer 1997).  Since ASCA can not separate the lines from the
continuum, we also test the method of not setting $\it{\Gamma}$(soft) equal to 
$\it{\Gamma}$(hard).  For this case, we determine the intrinsic luminosity 
in the $0.5-2$ keV band by assuming a power-law model with $\it{\Gamma}$=1.7,
extrapolating this model to lower energies and calculating the flux.  The 
conversion factor determined by extrapolating the luminosities is deduced to 
be 0.55, which allows us to obtain the scattering 
efficiency.  We note that these scattering efficiency is consistent with 
those deduced from the ratio of the normalizations within 20$\%$. 
The scattering efficiency results from an optical 
depth($\tau_{es}$) of a scattering medium multiplied by a covering factor 
of the scatterer($\Delta\Omega$).   
\begin{eqnarray}
(scattering~~efficiency) & = & \frac{L_{\rm{scat}}}{L_{\rm{intrinsic}}} \sim 
\frac{\Lsoft}{\Lints}, \nonumber \\
 & & \\
 & \sim & \tau_{es}\times\frac{\Delta\Omega}{4\pi} \nonumber
\end{eqnarray}

The $\tau_{es}$ is given by the column density ($N_{\rm{H}}^{\rm{scat}}$) in the 
scattering region multiplied by the Thomson
scattering cross section.  For the hard component, we infer
that the true AGN activity is greater than $\Lxa$. We define $B$ as the ratio 
 $\Lint/\Lxa$.  From eq.(3), $N_{\rm{H}}^{\rm{scat}}$ is calculated by the 
following relation,

\begin{equation}
N_{\rm{H}}^{\rm{scat}} \sim 1\times10^{24}~
(\frac{\frac{\Delta\Omega}{4\pi}}{0.133})^{-1}~(\frac{\frac{\Lsoft}{\Lxas}}{0.1})~B^{-1}~~~
  (\rm{H/cm^{2}}).
\end{equation}
 
The upper panel in Figure 7 shows the distribution of $\Lsoft/\Lxas$.
The points concentrate around 10$\%$. 
Taking the ratio to be 0.1, the column density in the scattering region is 
calculated to be 10$^{24} B^{-1}$ cm$^{-2}$, when the covering factor is 
0.133, which corresponds to the opening angle of the torus of 60 degrees.  
The analysis of 
the hard X-ray component supports the large-$B$ case. For example,
for $B$=10, the column is estimated to be 10$^{23}$ cm$^{-2}$, which is less 
than or
comparable to those of our line of sight.  The optical depth ($\tau_{es}$) of
the scatterer is deduced to be $\sim$0.06, which is similar to those used in 
the analysis of \PBLsey~(e.g. Miller \& Goodrich 1990).

We can also estimate the location of the scattering medium.  From eq. (4), 
the column density in the scattering region is 10$^{24} B^{-1}$ cm$^{-2}$. 
If the scatterer were cold, the 
soft X-rays would be absorbed by carbon and oxygen.
We do not observe a low-energy cut off and so carbon and oxygen must be 
ionized in the scatterer.
Kallman \& McCray (1982) pointed out that for log $\xi > 2$, the 
population of fully ionized carbon and oxygen is $\sim$100 $\%$ and 
$>$30$\%$, respectively, where $\xi$ is the ionization parameter with the 
relation of  $\xi$=$L$/($n_{\rm{e}}R^{2}$).  In log$\xi >$ 2, the electron temperature 
is greater than 10$^5$  K (Kallman \& McCray 1982), which is 
roughly consistent with the prediction with spectropolarimetry 
(e.g. Tran 1995b).  The ionization parameter,$\xi$, is calculated from the
observed value (see Eq. (5)), i.e. X-ray luminosity and column density of the 
scatterer.

\begin{equation}
\xi = 30~(\frac{\Lxas}{10^{43} \rm{ergs s^{-1}}})~(\frac{N_{\rm{e}}}
{10^{24} \rm{cm^{-2}}})^{-1}(\frac{R}{1\rm{pc}})^{-1}~ B.
\end{equation}

To obtain a large $\xi$, the scatterer should be located within 0.3 pc
for $B$=1 and within 30 pc for $B$=10.  For the most probable dusty torus
model for Seyfert nuclei (e.g. Murayama, Mouri, \& Taniguchi 2000), the 
full height of the torus is estimated to be $0.3-1.5$ pc.  Assuming $B$=1, the 
scatterer is located within the torus, while for $B$=10, the scatterer is 
located outside the torus.  Since large $B$ is more likely from the analysis 
of the hard component, the scattering region should be larger 
than that of the dusty torus.

Variability restricts the size of an emitting region.
For Mrk 3, long-term monitoring for 10 years shows that 
although the hard component changed by a factor of 6 during 3 years,
no decrease of the soft component was seen.
If most of the soft X-rays from Mrk 3 are scattered, the flux 
of the soft component should decrease within the light travel time 
from the nucleus to the scattering region.  The large time lag
indicates that the distance of the 
scatterer from the nucleus is comparable to 6 light years ($\sim$ 2 pc).
This result supports a large scattering region.  Continuous monitoring of
Mrk 3 in the X-ray band is crucial to verify the location of the scattering 
region.

\subsubsection{Comparison with Other Seyfert 2 Galaxies}

Ueno (1995) estimate a ratio of $\Lsoft/\Lxas$ for Seyfert 2 galaxies 
without reports of polarized broad lines using the ratio of the normalization
of their soft and hard components (Table 4).
For comparison, we show a histogram of $\Lsoft/\Lxas$ for
Seyfert 2 galaxies with and without reports of polarized broad lines
(Figure 7).  For typical Seyfert 2s, the 
ratios tend to be less than those of \PBLsey.   
This is explained either by a stronger soft component 
or a weaker hard component in \PBLsey.
To examine which applies, we compare $\Lxa$-$\Lir$  
for both types, assuming that $\Lir$ represents the real nuclear
activities because their infrared colors are warm except NGC 1068 (as 
discussed above).  
Figure 8a shows a color-color diagram of $\Lsoft/\Lxas$ and 
$\Lxa/\Lir$. Closed circles and open circles represent PBL Seyfert 2 and 
other Seyfert 2 galaxies, respectively.  When the absorption 
correction is applied, there is a clear separation of the two types of 
Seyfert 2 galaxies. For the reflection correction, using $\Lints$ in place of 
$\Lxas$ (Figure 8b), the data
are all concentrated in the same region, since PBL Seyfert 2s have larger 
$\Lxas/\Lints$ ratios than those of other Seyfert 2 galaxies. 
This result indicates that the distributions of $\Lsoft/\Lints$
in Figure 7 are due to underestimation of their nuclear activities, and 
that the scattering efficiencies of \PBLsey~are similar to 
those of other Seyfert 2 galaxies, which is a few $\%$.    

\subsubsection{Interpretation on PBL Seyfert Galaxies}

We find that the ratios of $\Lxa/\Lir$ of \PBLsey~are different from those 
of other Seyfert 2 galaxies.  We propose that the difference is caused
by viewing angle.  First, about half of our \PBLsey~were 
selected among Seyfert 2 galaxies with high polarization degrees as 
mentioned by Miller \& Goodrich (1990).  
If the scattering region is larger than the torus, 
most of the scattering region is not obscured.  The polarization degree
($\it{P}$) is described by the function of the viewing angle ($\it{i}$), 
$P = \sin^{2}i/(2\alpha+\sin^{2}i)$, where $\alpha$ is the value derived
from cone angle of the torus (see Miller \& Goodrich 1990 in detail).
The difference in the amount of polarization 
is explained as other Seyfert 2 galaxies having smaller viewing 
angles than \PBLsey. 
Next, the contribution of the Compton reflection component in the $2-10$ keV band
probably depends on the viewing angle, because most Seyfert 2 galaxies with 
pure reflection in the $2-10$ keV band have an H$_{2}$O maser line which is 
explained by their edge-on geometry (e.g. Matt et al. 1999 for the circinus 
galaxy, Ueno et al. for NGC 1068, Iwasawa et al.  for NGC 4945). Furthermore,
Bassani et al. (1999) find an anti-correlation between iron line equivalent 
width and $F_{\rm{X}}$/$F_{\rm{[OIII]}}$ for a large sample including 
Seyfert 1 galaxies and Seyfert 2 galaxies with the H$_{2}$O masers.

We suggest that the X-ray properties can be 
smoothly connected from Seyfert 1 galaxies to Seyfert 2 galaxies with 
the H$_{2}$O masers lying between ``typical'' Seyfert 2 
galaxies and \PBLsey. 
We conclude that the separation seen in 
Figure 8a results from smaller $\Lxa/\Lir$ for \PBLsey,
and that $\Lxa/\Lir$ decreases as the viewing angle increases.
Our picture is consistent with the fact that the viewing angle of Mrk 348 is 
smaller than the other \PBLsey~(Tran 1995b).  
Heisler, Lumsden, $\&$ Bailey (1997) proposed that \PBLsey\ have smaller 
viewing angle than other Seyfert 2 galaxies from the analysis of their 
extinction,$\it{E(B-V)}$. However X-ray properties suggest that the scattering
region is larger than the torus, and that \PBLsey\ have larger viewing angle
than other Seyfert 2 galaxies.

We consider two possibilities to explain the viewing angle dependence of  
$\Lxa/\Lir$; 1) anisotropic radiation from the Seyfert 2 nucleus due to 
X-ray beaming and 2) the existence of thick, clumpy clouds that block a 
considerable fraction of X-ray photons in the $2-10$ keV band (Figure 9).  
In the former model, anisotropic radiation can result from a jet structure found 
in radio band (Ulvestad \& Wilson 1984).  
The latter model is referred to as the dual absorption model 
(Weaver et al. 1994), in which absorption
columns are $N_{\rm{H}}\sim$10$^{23}$ and $>10^{24}$ for \PBLsey. 
Since this model indicates the some fraction of the X-rays are not absorbed by 
a thick matter, the thick matter should be smaller than the size of X-ray 
emitting 
region. In NGC 4258, it is found that H$_{2}$O vapor masing clouds are 
very small, thick clouds located in its central nuclear region 
(Miyoshi et al. 1995).  Thus the masing cloud is one candidate of the 
thick matter found in our observations.  Although we have proposed two 
possibilities, we have not obtained conclusive result on this issue, because 
we only obtained the $0.5-10$ keV spectra with poor statistics.  
This is a subject for future missions with
large effective area and high energy resolution, such as XMM, Astro-E, 
Constellation-X, and XEUS.

\section{CONCLUSION}

We analyze the X-ray data for Seyfert 2 galaxies which have polarized broad 
lines in the optical band. 

All galaxies show obscured nuclear activities with $2-10$ keV
absorption-corrected  
luminosities of $10^{42}-10^{43}$ \ulum.  However, absorption corrected
 X-ray luminosities ($\Lxa$) are less than expected based on predictions from 
$\Lir$, $\Lhbeta$ using relations found in Seyfert 1 
galaxies.  We examine whether the small X-ray luminosities are real or not,
and find that this can be explained if the effects of Compton reflection 
rather than absorption dominate the $2-10$ keV spectra.
In this case, the intrinsic X-ray luminosities are consistent with the 
higher luminosities expected from the amount of reflection. 

Soft X-ray emission is detected in all galaxies.  The X-ray spectrum and 
correlation of $\Lir$-$\Lsoft$ and $\Lhbetascat$-$\Lsoft$ 
reveal that most X-rays in the soft component are scattered light by a warm 
scatterer.  The typical scattering efficiency derived from the ratio 
of the soft to 
apparent hard band luminosity is about 10$\%$; 
larger than that of Seyfert 2 galaxies without  
polarized broad lines.  This discrepancy can result from an 
underestimation of intrinsic activity.
The scattering efficiencies deduced from real activities are similar 
to those of other Seyfert 2 galaxies, which are a few $\%$.  The small 
scattering efficiency supports the existence of a large scattering region,
as opposed to the compact scattering region proposed by Hisler, Lumsden, \& 
Bailey (1997).

We suggest that \PBLsey~have larger viewing angles and are seen  
more edge-on than 
other Seyfert 2 galaxies.   For this case, the polarization and the 
contribution of scattered light in the hard component for Seyfert galaxies
are naturally explained. We consider two possibilities to explain the 
viewing angle dependence of  $\Lxa/\Lir$; 1) anisotropic radiation and 
2) complex absorption by thick, clumpy clouds.  The ASCA data do not 
provide conclusive results on this issue. This is a subject to 
future X-ray missions.

\acknowledgements

The authors acknowledge all member of ASCA team.
This research has made use of the NASA/IPAC Extragalactic Database (NED) 
operated by the Jet Population Laboratory, and of the ASCA archive data base
maintained by the ASCA Guest observatory facility at NASA/GSFC.  
This work is supported by Japan Science and Technology Corporation. 
YT was financially supported in part by the Ministry of Education, Science, 
and Culture (Nos. 10044052, and 10304013).

\newpage

\begin{deluxetable}{lcccc}
\tablenum{1}
\tablewidth{0pt}
\tablecaption{observation log}
\tablehead{
\colhead{Target Name}  &  \colhead{Obs.Date} & \colhead{Exp. Time} &
\colhead{count rate(SIS)} & \colhead{hardness ratio(2-10/0.5-2)} }
\startdata
Was 49b  &   95. 5.22  &  43k  &  0.009$\pm$0.001 cts/s &  1.5$\pm$ 0.2   \nl 

Mrk 348  &   95.08.04  &  28k  &  0.027$\pm$0.001 cts/s &  8.0$\pm$ 1.0   \nl 

Mrk 1210 &   95.10.18  &  14k  &  0.016$\pm$0.002 cts/s &  1.1$\pm$ 0.1   \nl
         &   95.11.12  &   6k  &  0.017$\pm$0.006 cts/s &  1.3$\pm$ 0.4   \nl

NGC 7212 &   95.11.13  &  16k  &  0.011$\pm$0.001 cts/s &  1.2$\pm$ 0.2   \nl
         &   95.11.19  &  13k  &  0.009$\pm$0.002 cts/s &  0.94$\pm$0.2   \nl

Mrk 477  &   95.12.04  &  30k  &  0.011$\pm$0.001 cts/s &  1.6$\pm$ 0.1   \nl 
         &   95.12.05  &  20k  &  0.010$\pm$0.002 cts/s &  1.4$\pm$ 0.2   \nl

Mrk 3    &   96.10.26  &  32k  &  0.047$\pm$0.001 cts/s &  1.04$\pm$ 0.04 \nl 

\enddata
\end{deluxetable}

\begin{deluxetable}{lccccccccc}
\tiny
\tablenum{2}
\tablewidth{0pt}
\tablecaption{Fitting results for the ASCA spectra.}
\tablehead{
\colhead{}             & \colhead{} & \multicolumn{3}{c}{Soft Component} & 
                         \multicolumn{4}{c}{Hard Component} &
                         \colhead{} \\
\cline{3-5} \cline{6-9}  \\
\colhead{Target}  &  \colhead{ID$^{a}$} & \colhead{$N_{\rm{H0}}$} & \colhead{$\Gamma$/kT} &
\colhead{Z}  &  \colhead{$N_{\rm{H1}}$} & \colhead{$\Gamma$}  &
\colhead{Center Energy} & \colhead{EW}  &  \colhead{$\chi^{2}$ (d.o.f.)} \\
\colhead{}  &  \colhead{} & \colhead{($\times$10$^{22}$ cm$^{-2}$)} & \colhead{--/(keV)} &
\colhead{} & \colhead{($\times$10$^{22}$ cm$^{-2}$)}   & \colhead{} &
\colhead{(keV)} & \colhead{(eV)} & \colhead{} \\
}
\startdata
Was 49b  & 1 &  0.85($<$1.2)  & 8.9(2.3--15)  & ----  & 0.16($<$0.56) & 0.54(0.25--0.9)   &  5.7(5.4--6.1) & 370(200--500) &  70.0(62) \nl
         & 2 &  0.0163(f)     & 2.3(0.5--3.0) & ----  & 6.3(4.5--46)  &   1.7(f)          &  6.02(f)       & 620(370--870) &  91.2(65) \nl
         & 3 & 0.56( 0.2--1.14) & 7.2(4--10)  & ----  &     ----      & 2.8(2.48-- 3.12 ) &  6.02(f)       & 200($<$350)   &  71.4(64) \nl
         & 4 & 0.0163(f) & 0.72(0.4--1.1) & 0.02($<$0.14)  & 1.7(0.7--3.2) & 0.6(0.3--0.9) &  6.02(f)      & 360(180--540) &  72.0(63) \nl

Mrk 348  & 1 & 0.00($<$0.24) & 1.1(0.5--2.3) & ----   & 16(13--20)    & 1.69(1.30-- 2.05) & 6.27(6.20--6.35) & 215(140--300) &  117.4(116) \nl
         & 2 & 0.056(f)      & 1.3(0.7--1.9) & ----   & 16(15--18)    &  1.7 (f)          & 6.27(6.20--6.35) & 212(140--280) &  118.6(119) \nl
         & 3 & 0.056 (f)     & -3            & ----   &   ----        &  1.6              &  4.9             & 200           &  242(117)   \nl
         & 4 & 0.056 (f)     & 64( 6$>$ )    &  1(f)  & 16(15--18)    &  1.64(1.30--2.00) & 6.27(6.20--6.35) & 215(140--300) &  117.6(117) \nl

Mrk 1210 & 1 & 0.83(0.16--1.4) & 7.3(3.4--10.7) & ----  & 0.00($<$1.7) & -0.14(-0.45-- +0.28) & 6.29(6.24--6.60) & 780(510--1020) &  33.2(35) \nl
         & 2 & 0.038(f)      & 2.34(1.90--2.80) & ----  & 21.1(15--30) &  1.7 (f)             & 6.315(f)         & 760(430--990)  &  50.1(38) \nl
         & 3 & 0.55(0.1--1.1) & 5.6(3.2--8.4)   & ----  &  ----        & 1.95(1.55--2.40)     & 6.315(f)         & 460(250--670)  &  32.9(36) \nl
         & 4 & 0.038(f) & 0.86(0.55--1.2) & 0.03($<$0.1)  & 3.7($<$7.5) & 0.08(-0.5--0.5)     & 6.315(f)         & 710(440--950)  &  34.8(36) \nl

NGC 7212 & 1 & 0.0($<$0.2) & 1.37(0.8--4.4) & ----  & 23(0--200)  & 0.47(-1.5--10)  & 6.18(---)$^a$ & 520(---)      &  30.2(34) \nl
         & 2 & 0.0533(f)   & 1.41(1.0--1.8) & ----  & 45(20--100) & 1.7(f)          & 6.24(f)       & 480($<$1500)  &  32.4(37) \nl
         & 3 & 0.0533(f)   & 1.65(0.2--2.3) & ----  &   ----      & 1.87(1.35--2.7) & 6.24(f)       & 340(20--700)  &  30.1(37) \nl
         & 4 & 0.0533(f)   & 12.0(4.5$<$)   & 1(f)  & 25($<$95)   & 0.77(-1.2--5.7) & 6.24 (f)      & 530(60--1200) &  32.1(36) \nl

Mrk 477 & 1 &  0.0($<$0.6)  & 2.3(1.6--6.5)   & ----  & 4.7($<$13) & -0.02(-0.6-- +0.6) & 6.24(6.27--6.39) & 500(350--650) &  71.1(70) \nl
        & 2 &  0.0127(f)    & 1.67(1.25--2.1) & ----  & 30(24--42) & 1.7(f)             & 6.17(f)          & 300(150--450) &  88.3(73) \nl
        & 3 & 0.01($<$0.32) & 2.75(2.1--4.8)  & ----  &   ----     & 1.57(1.2--1.9 )    & 6.17(f)          & 230(80--380)  &  69.7(71) \nl
        & 4 & 0.0127(f)     & 7.5(4.8$<$)     & 1(f)  & 14(6--24)  & 0.3(-0.4-- +2.0)   & 6.17(f)          & 340(180--450) &  82.4(72) \nl

Mrk 3    & 1 & 0.04($<$0.09)   & 1.57(1.41--1.75) & ----           & 72(50--100) & 2.22(1.2--3.3)   & 6.29(6.27--6.34) & 700(520--1100) &  148(127)   \nl
         & 2 & 0.087(f)        & 1.72(1.60--1.82) & ----           & 57(48--68)  & 1.7(f)           & 6.315(f)         & 760(590--930)  &  190.9(130) \nl
         & 3 & 0.20(0.12-0.30) & 2.47(2.1--2.9)   & ----           &   ----      & 1.14(0.80--1.48) & 6.315(f)         & 730(610--840)  &  187.0(128) \nl
         & 4 & 0.087(f)        & 0.98(0.88--1.08) & 0.05(0.03--0.09)  & 0($<$0.7) & -0.23(-0.40-- -0.04) & 6.315(f)   & 1100(950--1200) &  224.3(129) \nl
\enddata
\tablecomments{ID: (1) absorbed power law plus another absorbed power law model, (2) absorbed power law with photon index of 1.7 plus another power law model, 
(3) Compton reflection plus absorbed power law model, For reflection models, the reflection mirror has an inclination of $\theta$=0$^{\circ}$ and a coverage of $\Omega$=2$\pi$, (4) absorbed power law plus RAYMOND plasma.  For Mrk 3, we included a narrow line at $\sim$0.9 keV in the soft component. A fixed parameter is denoted by (f).}
\tablenotetext{a}{Since we did not obtain a significant iron line in this model, its confidence region is too large.}
\end{deluxetable}

\begin{deluxetable}{lccccccccc}
\tablenum{3}
\tablewidth{0pt}
\tablecaption{Observed X-ray flux and luminosity$^{a}$}
\scriptsize
\tablehead{
\colhead{Target Name} & \colhead{redshift$^{b}$} & \multicolumn{5}{c}{X-ray flux (\uflux)} & 
\colhead{}             &  \multicolumn{2}{c}{X-ray luminosity (\ulum)} \\
\cline{3-7} \cline{9-10} \\
\colhead{} & \colhead{}&  \multicolumn{2}{c}{soft component} & 
\colhead{}             &  \multicolumn{2}{c}{hard component} &
\colhead{} &  \colhead{soft component} & \colhead{hard component}  \\
\cline{3-4} \cline{6-7} \\
\colhead{}  & \colhead{} & \colhead{0.5-2keV} & \colhead{2-10keV} & 
\colhead{} & \colhead{0.5-2keV} & \colhead{2-10keV} &  \colhead{} & 
\colhead{0.5-2keV} & \colhead{2-10keV} }
\startdata
Was 49b   & 0.0630 & 8.6$\times10^{-14}$ & 3.7$\times10^{-13}$ & & 5.9$\times10^{-17}$ & 6.3$\times10^{-13}$ & & 1.4$\times10^{42}$ & 1.0$\times10^{43}$  \nl
Mrk 348   & 0.0149 & 7.0$\times10^{-14}$ & 2.8$\times10^{-13}$ & & 2.1$\times10^{-16}$ & 4.8$\times10^{-12}$ & & 6.3$\times10^{40}$ & 4.3$\times10^{42}$  \nl
Mrk 1210  & 0.0135 & 1.8$\times10^{-13}$ & 1.8$\times10^{-13}$ & & 3.1$\times10^{-18}$ & 1.4$\times10^{-12}$ & & 1.3$\times10^{41}$ & 1.0$\times10^{42}$  \nl
NGC 7212  & 0.0260 & 1.0$\times10^{-13}$ & 3.3$\times10^{-13}$ & & 5.0$\times10^{-22}$ & 5.0$\times10^{-13}$ & & 2.8$\times10^{41}$ & 1.4$\times10^{42}$  \nl
Mrk 477   & 0.0378 & 1.0$\times10^{-13}$ & 2.0$\times10^{-13}$ & & 2.6$\times10^{-19}$ & 1.1$\times10^{-12}$ & & 5.8$\times10^{41}$ & 6.4$\times10^{42}$  \nl
Mrk 3     & 0.0135 & 5.8$\times10^{-13}$ & 1.3$\times10^{-12}$ & & 4.0$\times10^{-25}$ & 4.2$\times10^{-12}$ & & 4.3$\times10^{41}$ & 3.1$\times10^{42}$  \nl
\tablenotetext{a}{we listed the flux in the model 2.}
\tablenotetext{b}{Data quated from NASA Extragalactic Database (NED).}
\enddata
\end{deluxetable}

\begin{deluxetable}{lcccccc}
\tablenum{4}
\tablewidth{0pt}
\tablecaption{X-ray, infrared and H$_{\beta}$p luminosities for \PBLsey}
\scriptsize
\tablehead{
\colhead{Target Name}  &  \colhead{$\Lsoft$ $^a$} & \colhead{$\Lxa$ $^b$} & \colhead{$\Lir$} &
\colhead{log($F_{\rm{25\mu m}}$/$F_{\rm{60\mu m}}$)} & \colhead{$L_{\rm{H_{\beta}p}}$} & \colhead{Reference} }
\startdata
Was 49b   &  1.4$\times10^{42}$ & 2.1$\times10^{43}$  & 3.5$\times10^{44}$ &       & 3.1$\times10^{40}$ & 1  \nl
Mrk 348   &  7.3$\times10^{40}$ & 8.0$\times10^{42}$  & 5.5$\times10^{43}$ & -0.19 & 2.6$\times10^{39}$ & 1  \nl
Mrk 1210  &  1.4$\times10^{41}$ & 2.3$\times10^{42}$  & 5.7$\times10^{43}$ &  0.04 & 1.5$\times10^{39}$ & 1  \nl
NGC 7212  &  3.3$\times10^{41}$ & 4.9$\times10^{42}$  & 4.0$\times10^{44}$ & -0.57 & 3.5$\times10^{39}$ & 1  \nl
Mrk 477   &  6.3$\times10^{41}$ & 1.8$\times10^{43}$  & 3.8$\times10^{44}$ & -0.41 & 7.3$\times10^{39}$ & 1  \nl
Mrk 3     &  6.1$\times10^{41}$ & 7.0$\times10^{42}$  & 1.2$\times10^{44}$ & -0.11 & 9.1$\times10^{39}$ & 1  \nl
\nl
NGC 1068  &  3.9$\times10^{40}$ & 5$\times10^{41}$    & 5.2$\times10^{44}$ & -0.32 & 2.5$\times10^{39}$ & 2, 3 \nl
Mrk 463E  &  1.2$\times10^{42}$ & 8$\times10^{42}$    & 9.5$\times10^{44}$ & -0.14 & 3.6$\times10^{40}$ & 1, 4 \nl
NGC 7674  &  4.0$\times10^{41}$ & 2$\times10^{42}$    & 6.6$\times10^{44}$ & -0.47 & 4.8$\times10^{39}$ & 1, 5 \nl
\tablenotetext{a}{the 0.5--2keV luminosity for soft component.}
\tablenotetext{b}{the 2--10 keV absorption corrected luminosity for hard component.}
\tablerefs{
(1) Tran 1995a; (2) Antonucci \& Miller 1985; (3) Ueno et al. 1994;
(4) Ueno et al. 1996; (5) Malaguti et al. 1998
}
\enddata
\end{deluxetable}

\begin{deluxetable}{lcc}
\tablenum{5}
\tablewidth{0pt}
\tablecaption{intensity of 0.9 keV structure in model 2}
\tablehead{
\colhead{Target Name}  &  \colhead{center energy} & \colhead{equivalent width}  }
\startdata
Was 49b   &  0.85 keV (fixed) &  0 ($<$90)eV  \nl
Mrk 348   &  0.89 keV (fixed) & 80 ($<$240)eV  \nl
Mrk 1210  &  0.89 keV (fixed) & 35 ($<$700)eV  \nl
NGC 7212  &  0.88 keV (fixed) & 12 ($<$120)eV  \nl
Mrk 477   &  0.86 (0.82--0.90) keV & 85 (10--160)eV  \nl
Mrk 3     &  0.915 (0.89--0.93) keV & 140 (100--200) eV  \nl
\enddata
\end{deluxetable}

\begin{deluxetable}{lc}
\tablenum{6}
\tablewidth{0pt}
\tablecaption{The ratio of $\Lsoft/\Lxas$ for \PBLsey~and Other Seyfert 2
galaxies}
\tablehead{
\colhead{Target Name}  &  \colhead{$\Lsoft/\Lxas$} }
\startdata
Was 49b   &  0.12  \nl
Mrk 348   &  0.017 \nl
Mrk 1210  &  0.11  \nl
NGC 7212  &  0.12  \nl
Mrk 477   &  0.064 \nl
Mrk 3     &  0.16  \nl
Mrk 463e$^{a}$  &  0.273 \nl
NGC 7674$^{b}$   &  0.365  \nl
\nl
IC 5063$^{c}$   &  0.014 \nl
NGC 4388$^{c}$  &  0.050 \nl
NGC 4507$^{c}$  &  0.007 \nl
IRAS1319-164$^{c}$ &  0.053 \nl
TOL1351-375$^{c}$  &  0.065 \nl
NGC 7172$^{c}$  &  0.001 \nl

\enddata
\tablerefs{
(a) Ueno et al. 1996; (b) Malaguti et al. 1998; (c) Ueno 1995.
}
\end{deluxetable}

\newpage
\onecolumn
\begin{figure}
\plottwo{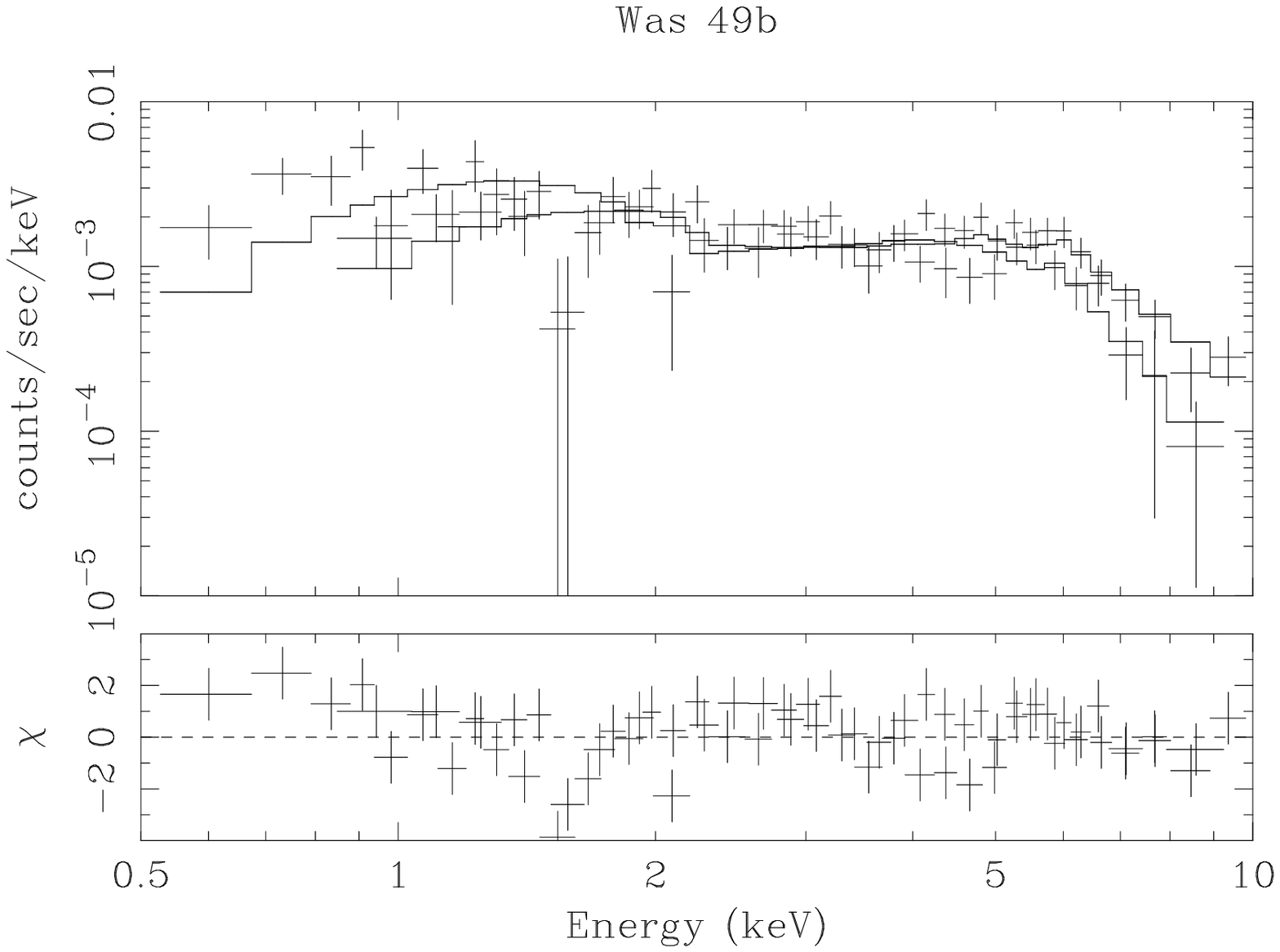}{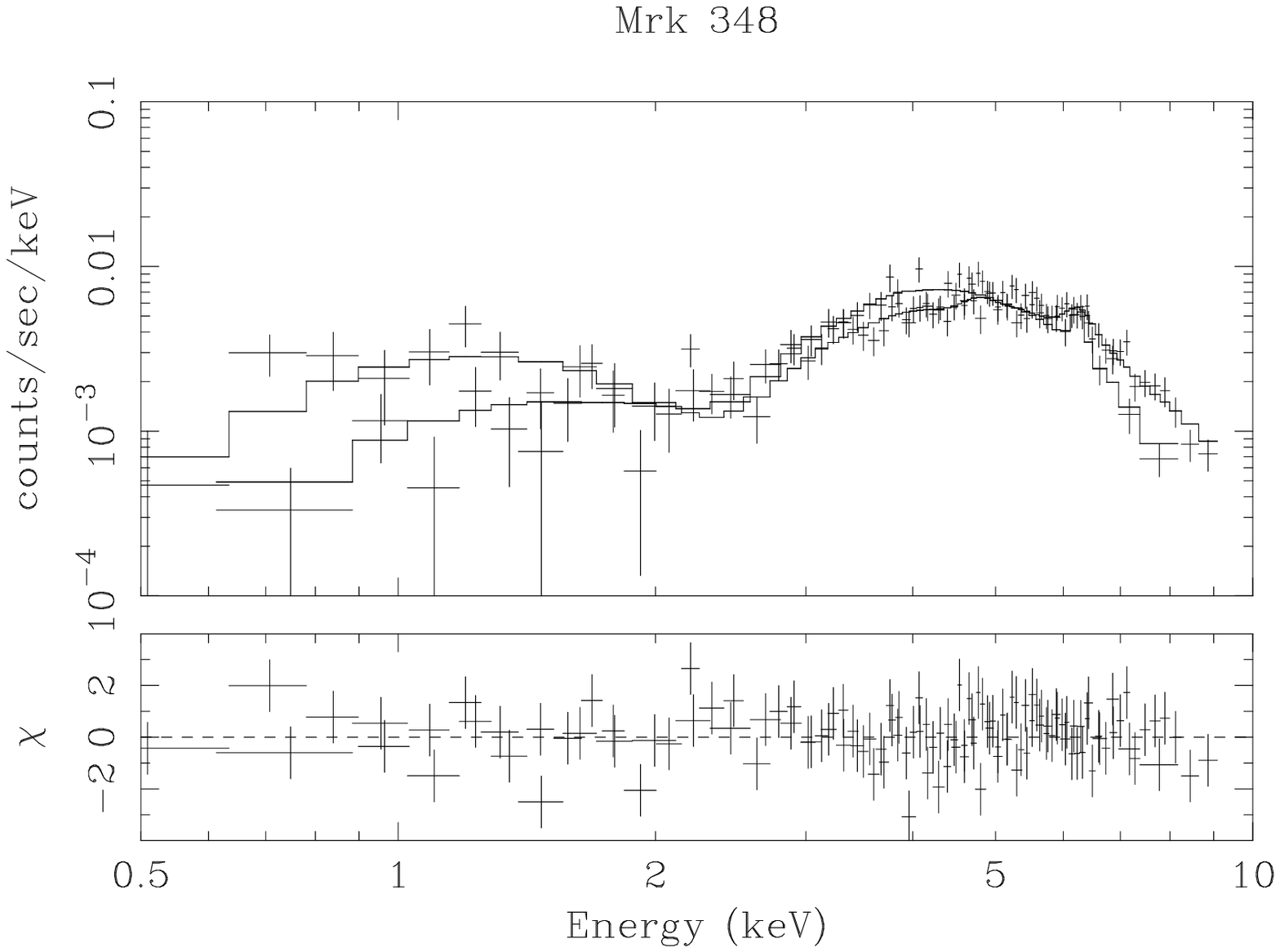}

\plottwo{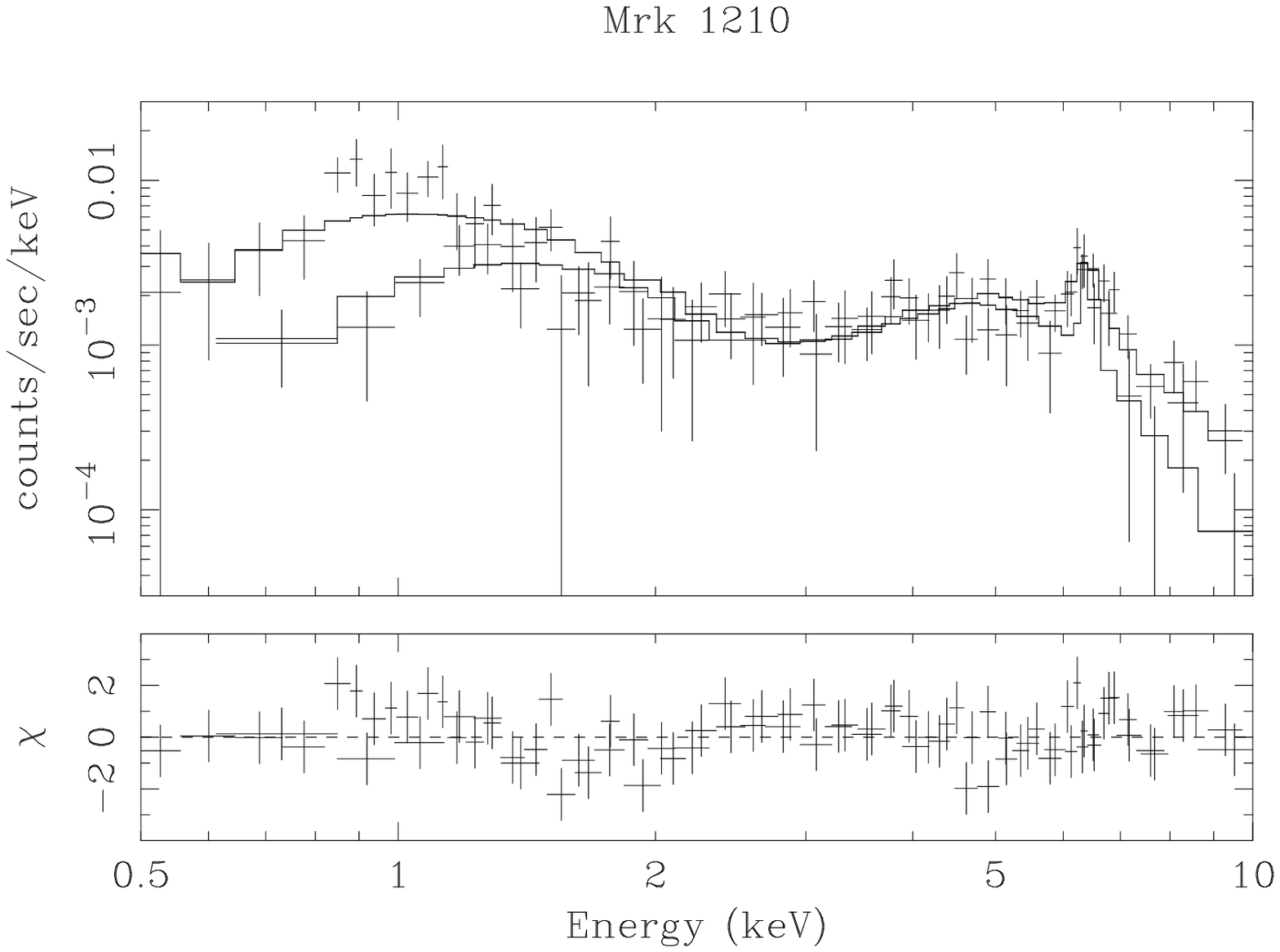}{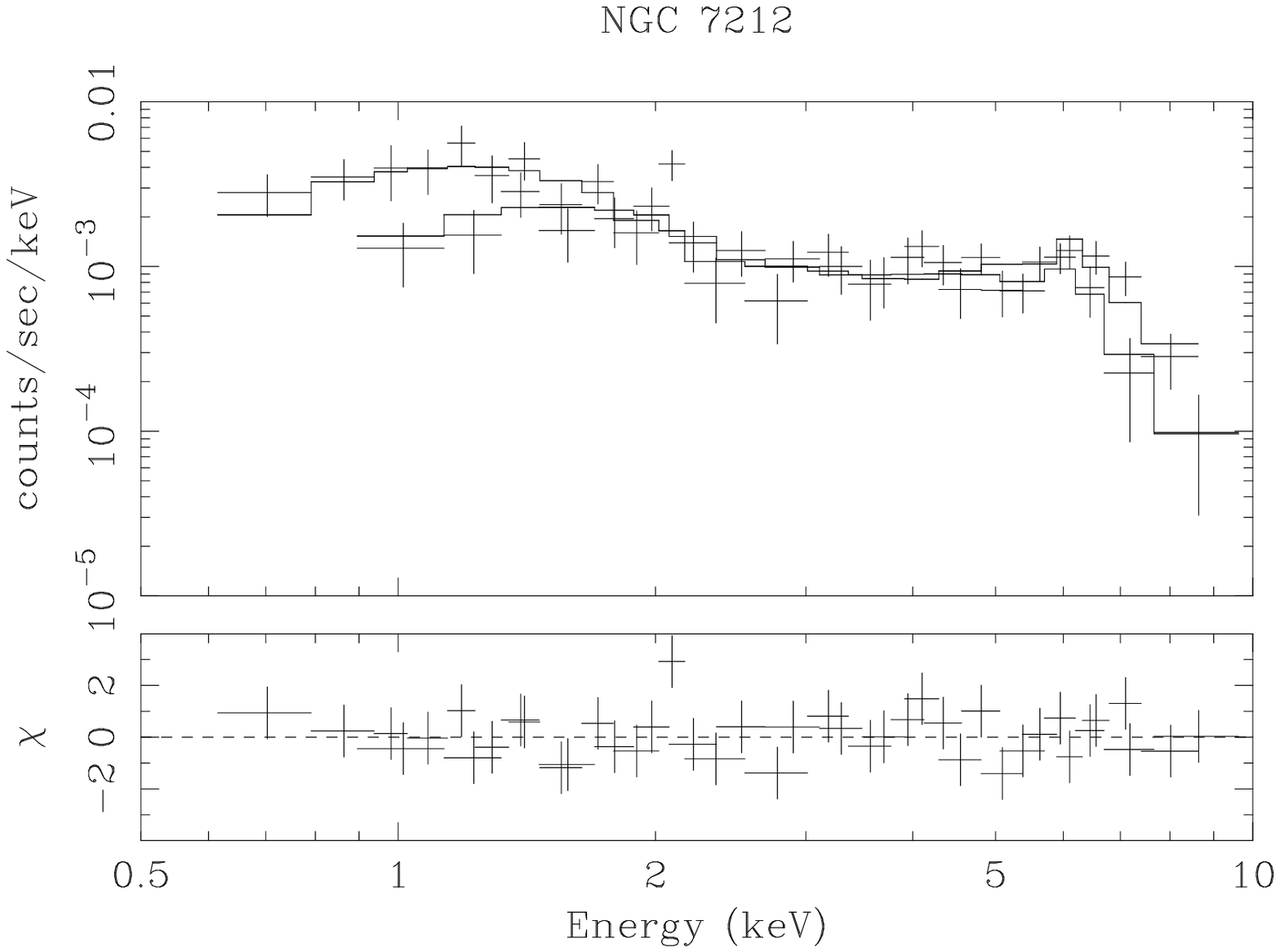}

\plottwo{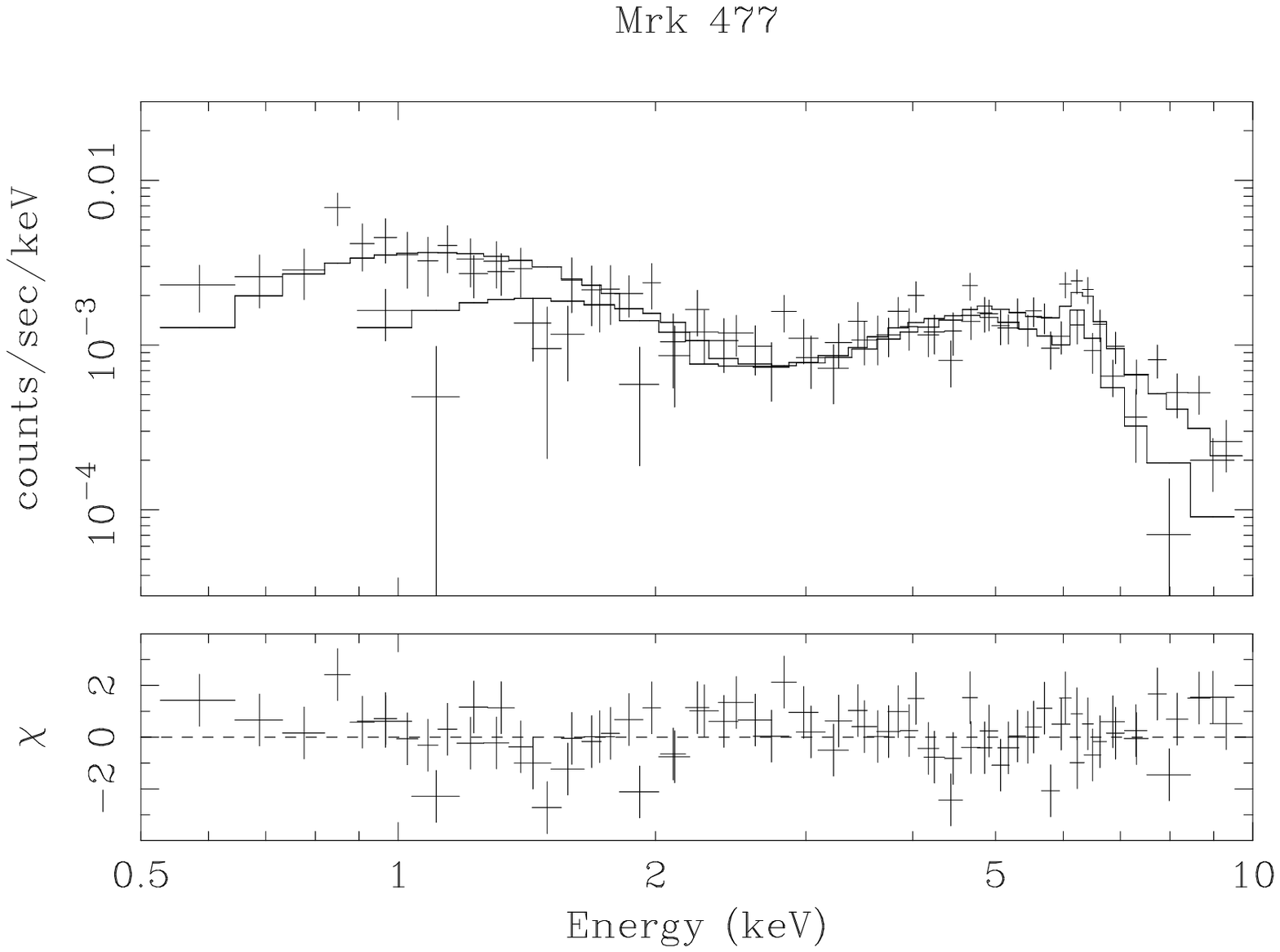}{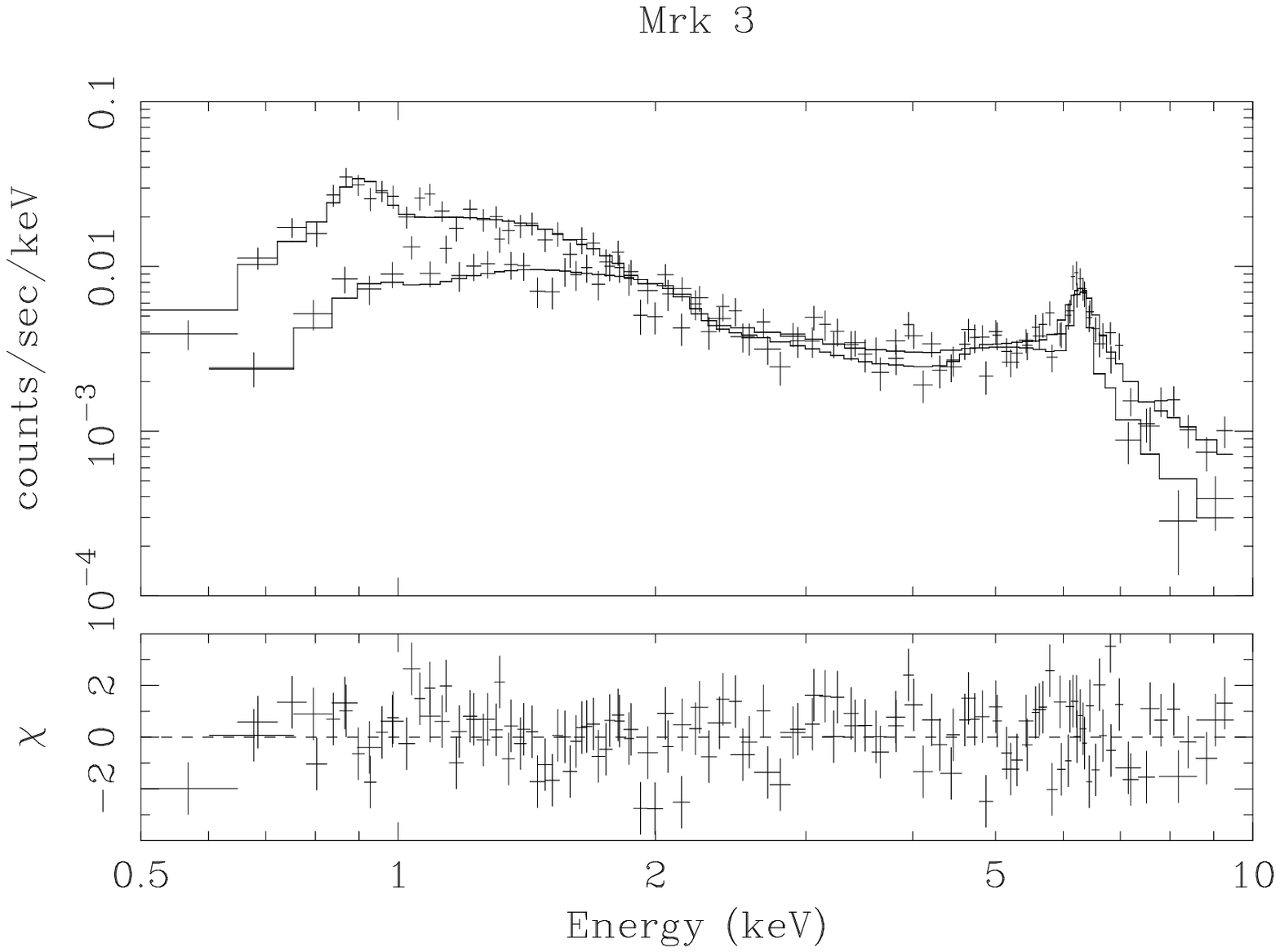}

\caption[figure1e.eps,figure1f.eps]{
ASCA SIS and GIS spectra of Seyfert 2 galaxies with optical polarized broad lines.
Histgrams show the best fit models in model 1. Please note 
that a narrow line at$\sim$0.9keV is included in the spectrum of Mrk 3.}
\label{figure:1}
\end{figure}

\begin{figure}
\epsscale{0.4}
\plotone{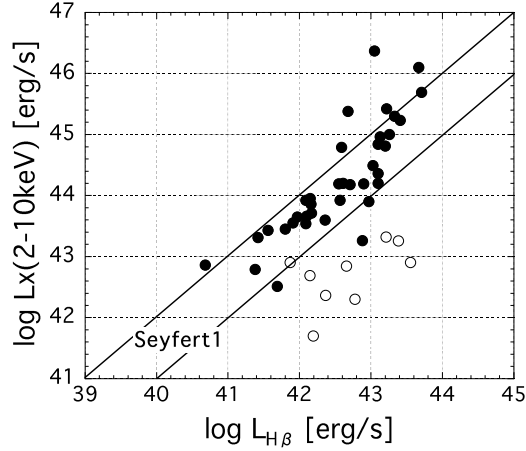}
\caption[figure2.eps]{
Absorption corrected X-ray luminosity ($\Lxa$) vs. broad H$\beta$ luminosity
($\Lhbeta$). Closed circles correspond to observations 
for Seyfert 1 galaxies and QSOs; open circles to estimations for Seyfert 2 galaxies 
with optical polarized broad lines. Two solid lines show $\Lxa/\Lhbeta$
=10 (lower) and 100 (upper).
}
\label{figure:2}
\end{figure}

\begin{figure}
\plottwo{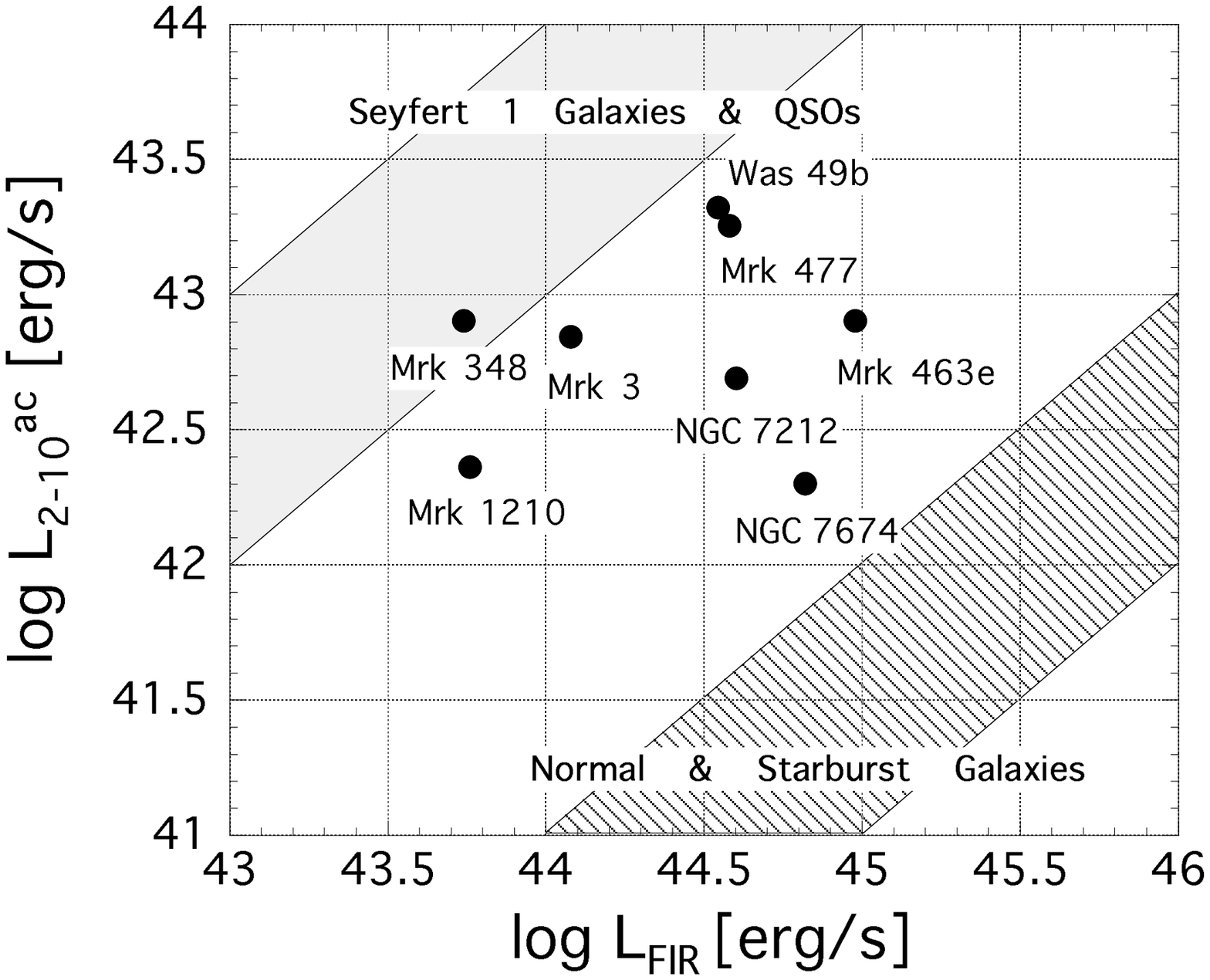}{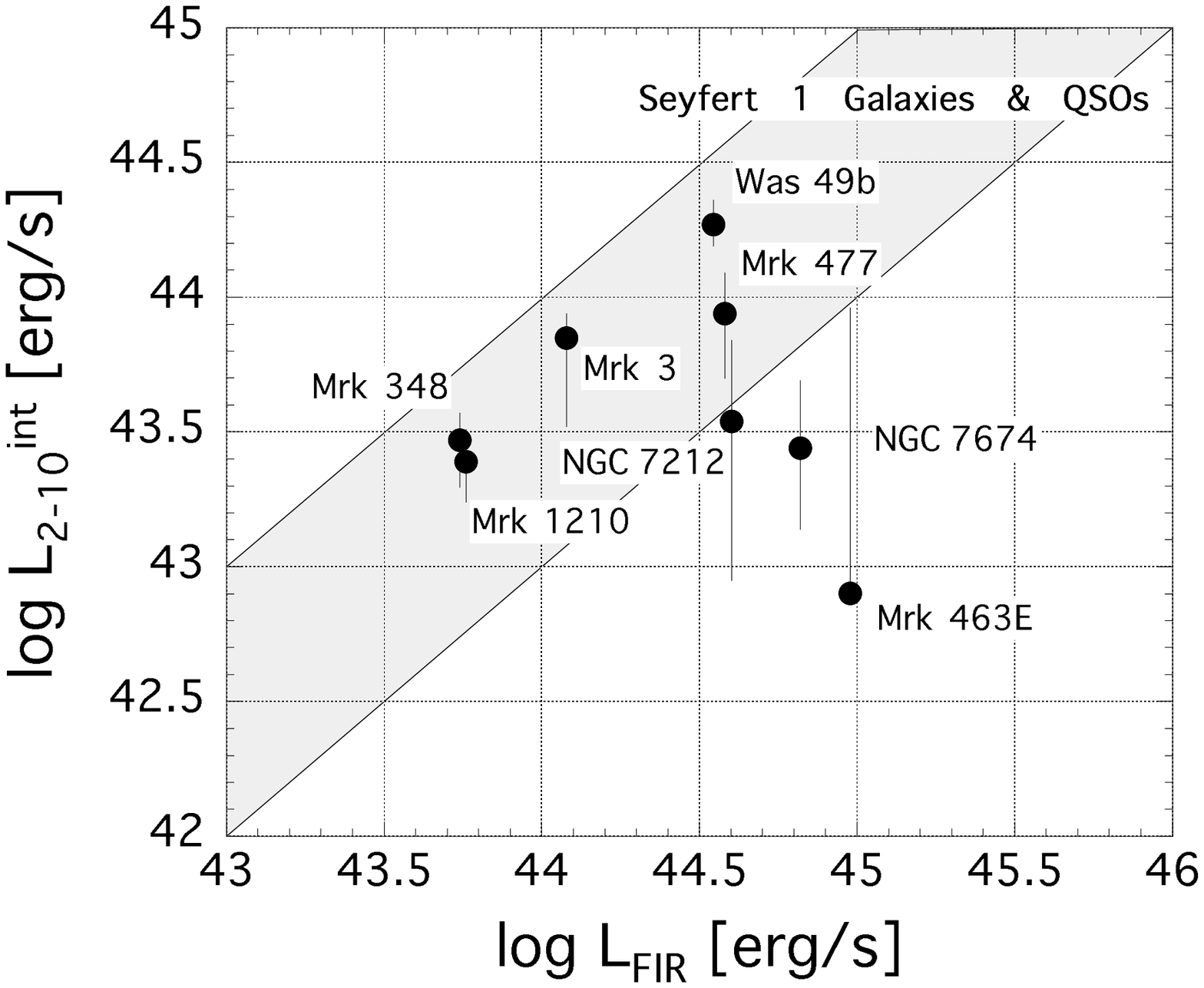}
\caption[figure3a.eps,figure3b.eps]{
Absorption corrected X-ray luminosity ($\Lxa$) vs. far-infrared 
luminosity ($\Lir$) (left), and 
intrinsic luminosity ($\Lint$) vs. far-infrared luminosity ($\Lir$)(right).
Intrinsic luminosities were estimated from eq.\ (2) in the text. The regions
for normal and starburst galaxies, and Seyfert 1
galaxies and QSOs are also presented in these figures.
}
\label{figure:3}
\end{figure}

\begin{figure}[h]
\epsscale{0.4}
\plotone{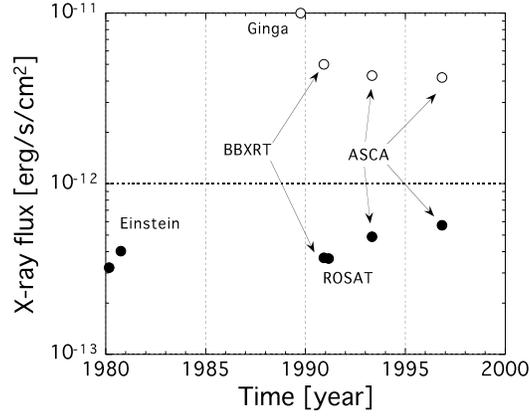}
\caption[figure4.eps]{
Long term monitoring of X-ray flux with various X-ray missions for Mrk 3. 
Closed circles 
and open circles show X-ray fluxes in the 0.5 -- 2keV and the 2 -- 10 keV, 
respectively.
}
\label{figure:4}
\end{figure}

\begin{figure}
\epsscale{0.4}
\plotone{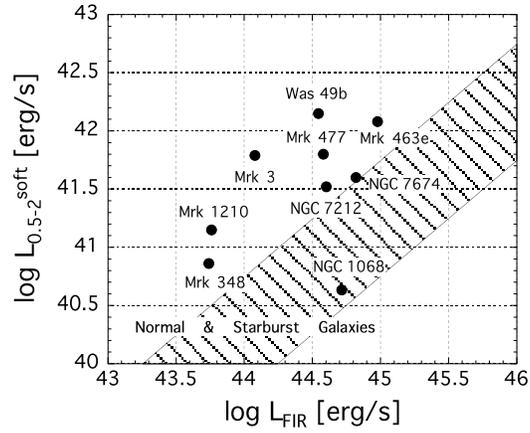}
\caption[figure5.eps]{The 0.5--2 keV luminosities ($\Lsoft$) vs. far-infrared luminosities 
($\Lir$). The 0.5--2keV luminosity for NGC 1068 is estimated by 
extrapolation of the scattered component seen in the 2--10 keV band. 
}
\label{figure:5}
\end{figure}

\begin{figure}
\epsscale{0.4}
\plotone{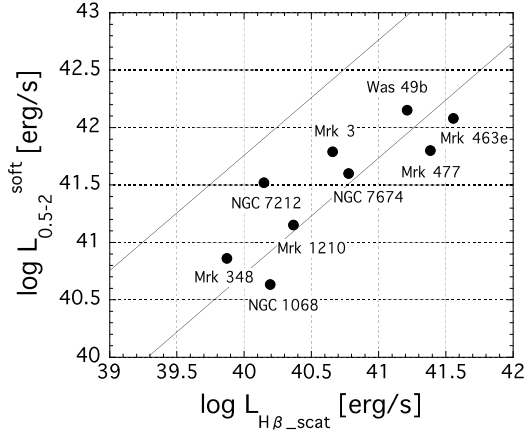}
\caption[figure6.eps]{The soft X-ray band luminosities ($\Lsoft$) vs. the scattered
broad H$\beta$ luminosities ($\Lhbetascat$).  These $\Lhbetascat$
are quoted from Tran 1995b and Miller \& Goodrich 1990. 
The two solid lines show $L_{\rm{2-10}}/\Lhbeta$=10 and 100 assuming a photon 
index of 1.7.
}
\label{figure:6}
\end{figure}

\begin{figure}[h]
\epsscale{0.4}
\plotone{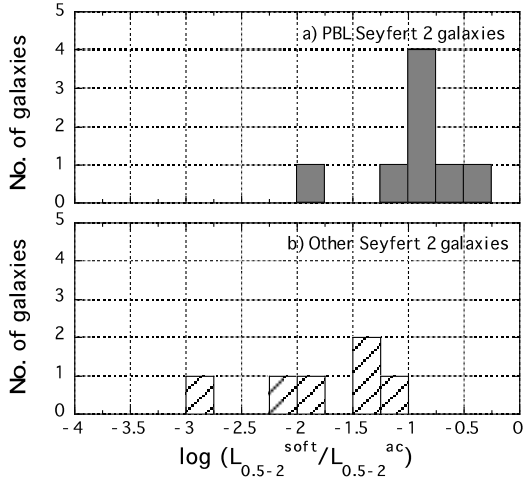}
\caption[figure7.eps]{
Distributions of $\Lsoft/\Lxas$ for PBL Seyfert 2 (upper) and for other 
Seyfert 2 galaxies (lower). The data for other Seyfert 2 galaxies are 
quated from Ueno (1995).
}
\label{figure:7}
\end{figure}

\begin{figure*}
\plottwo{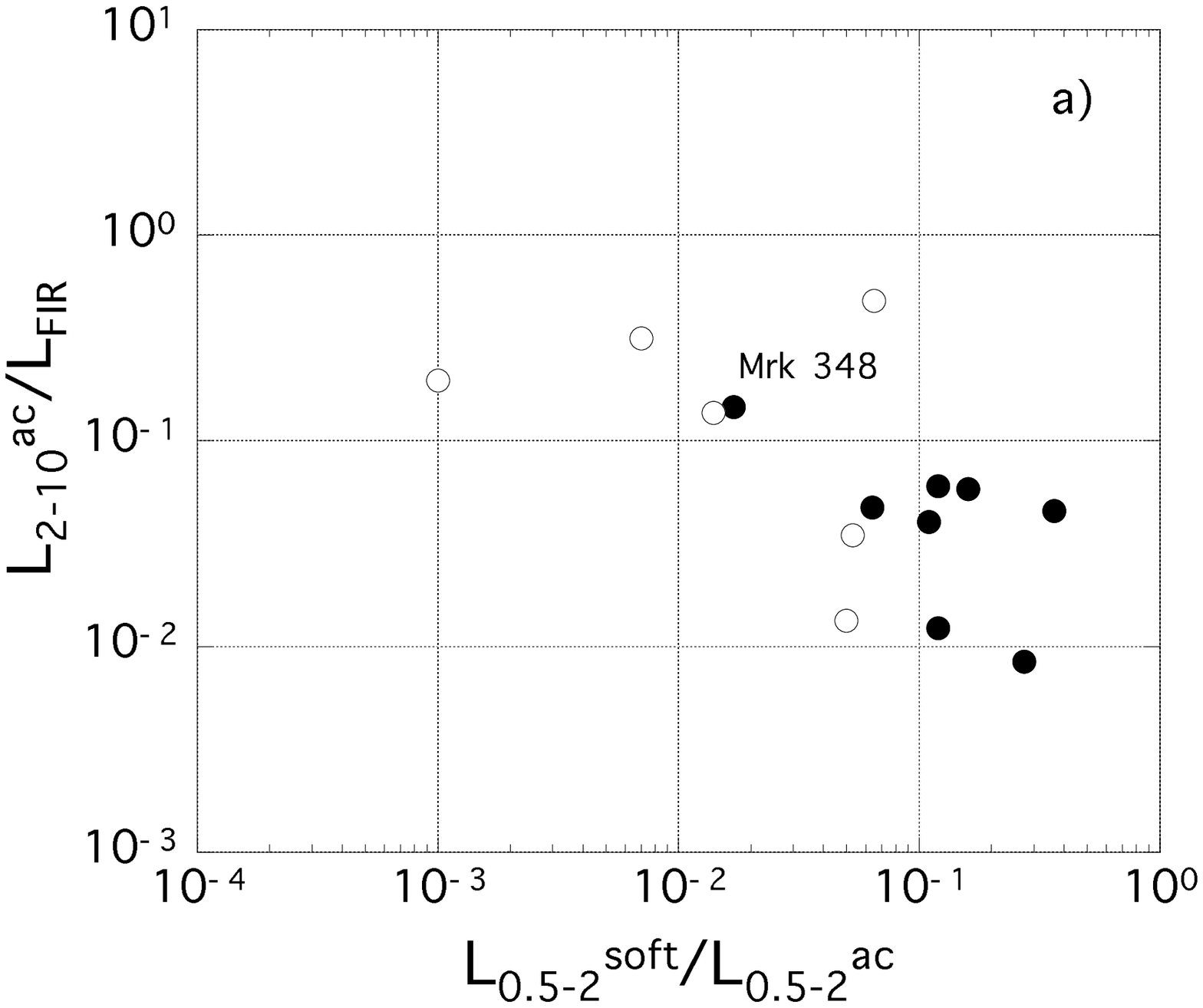}{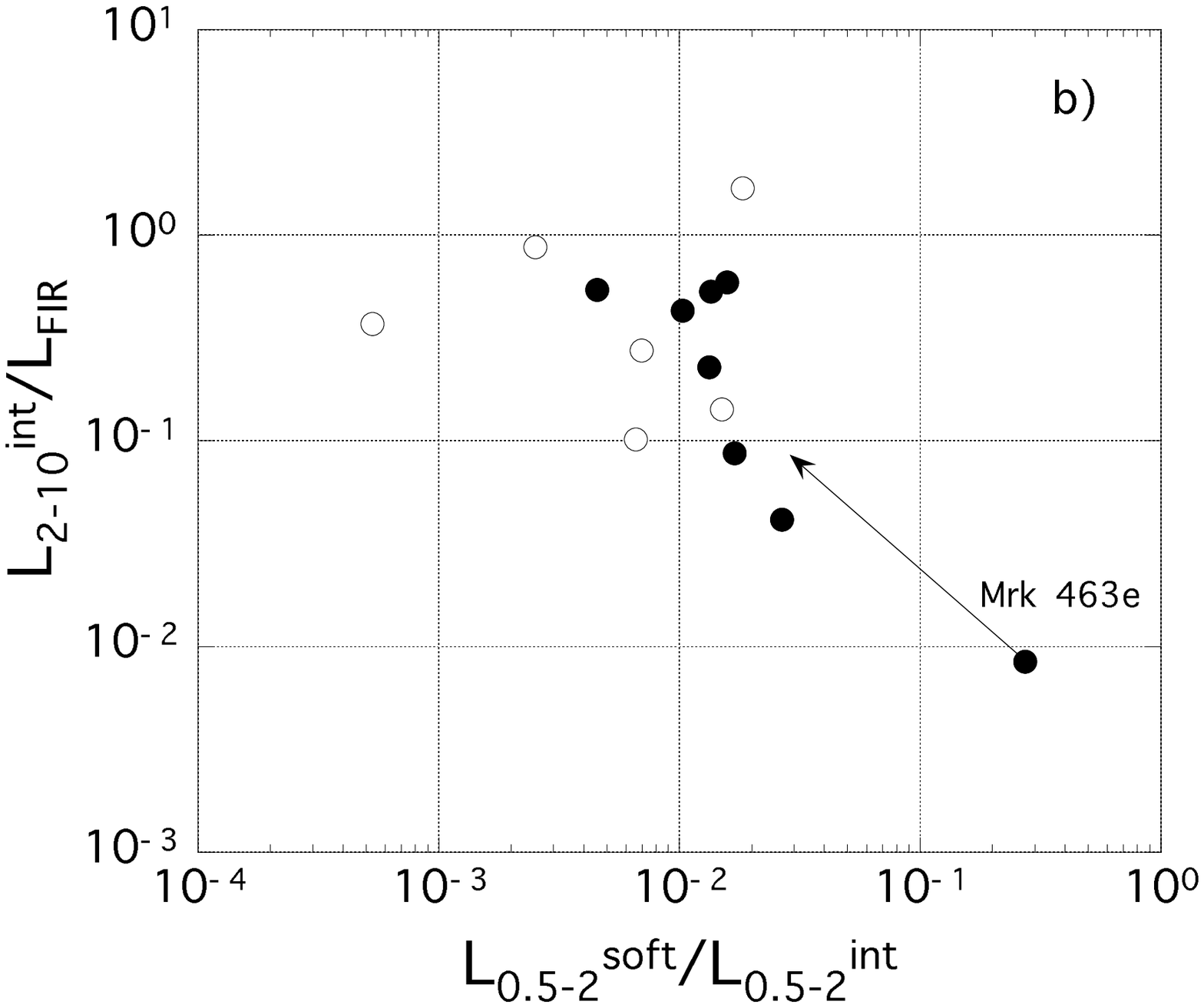}
\caption[figure8a.eps,figure8b.eps]{
 $\Lxa/\Lir$ vs $\Lsoft/\Lxas$ (left figure) and 
$\Lint/\Lir$ vs $\Lsoft/\Lints$ (right figure). Closed and open circles 
display  PBL Seyfert 2 and other Seyfert 2 galaxies, 
respectively. An arrow in figure 8b shows the error region of Mrk 463e.
}
\label{figure:8}
\end{figure*}

\begin{figure*}
\plottwo{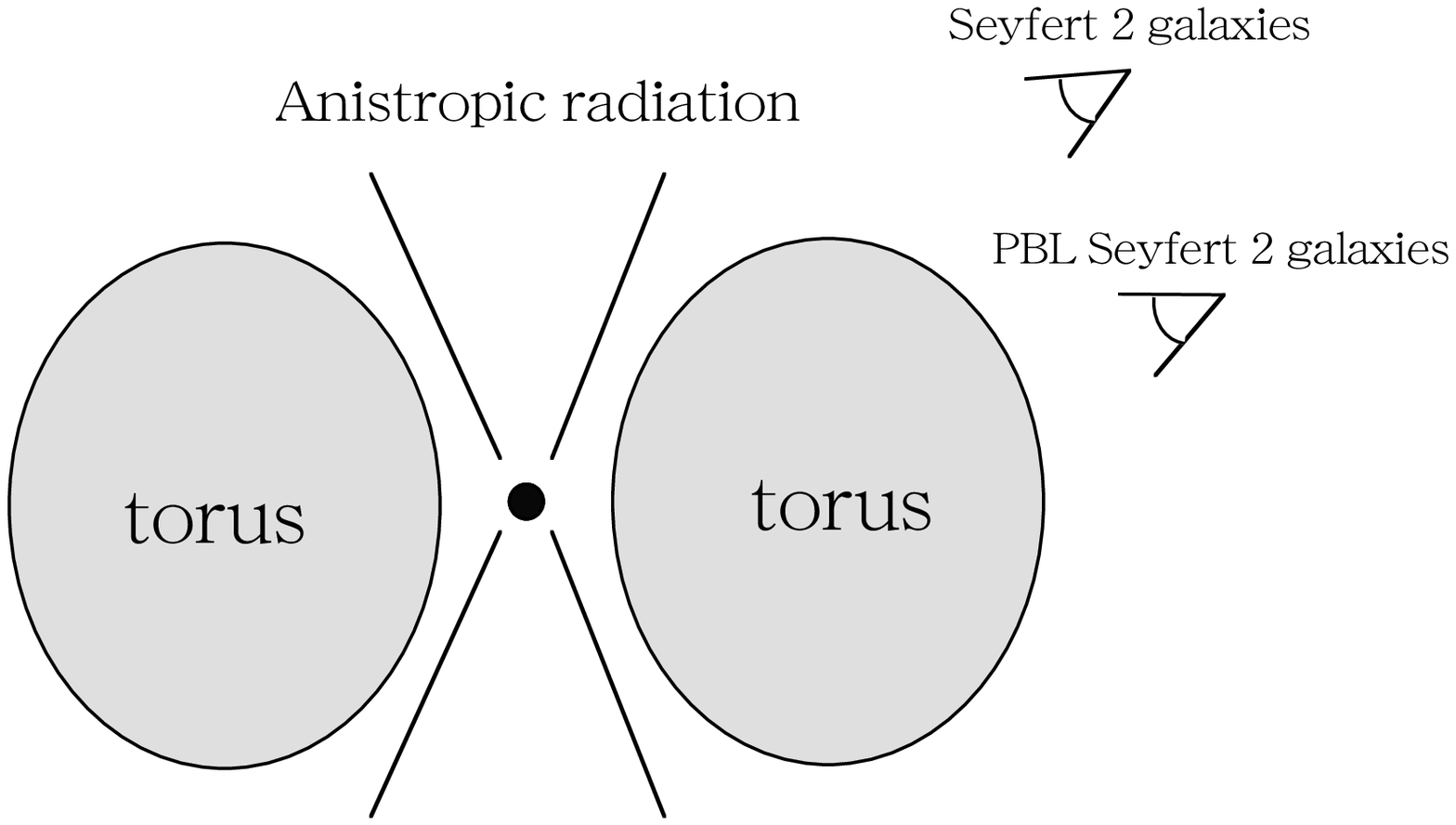}{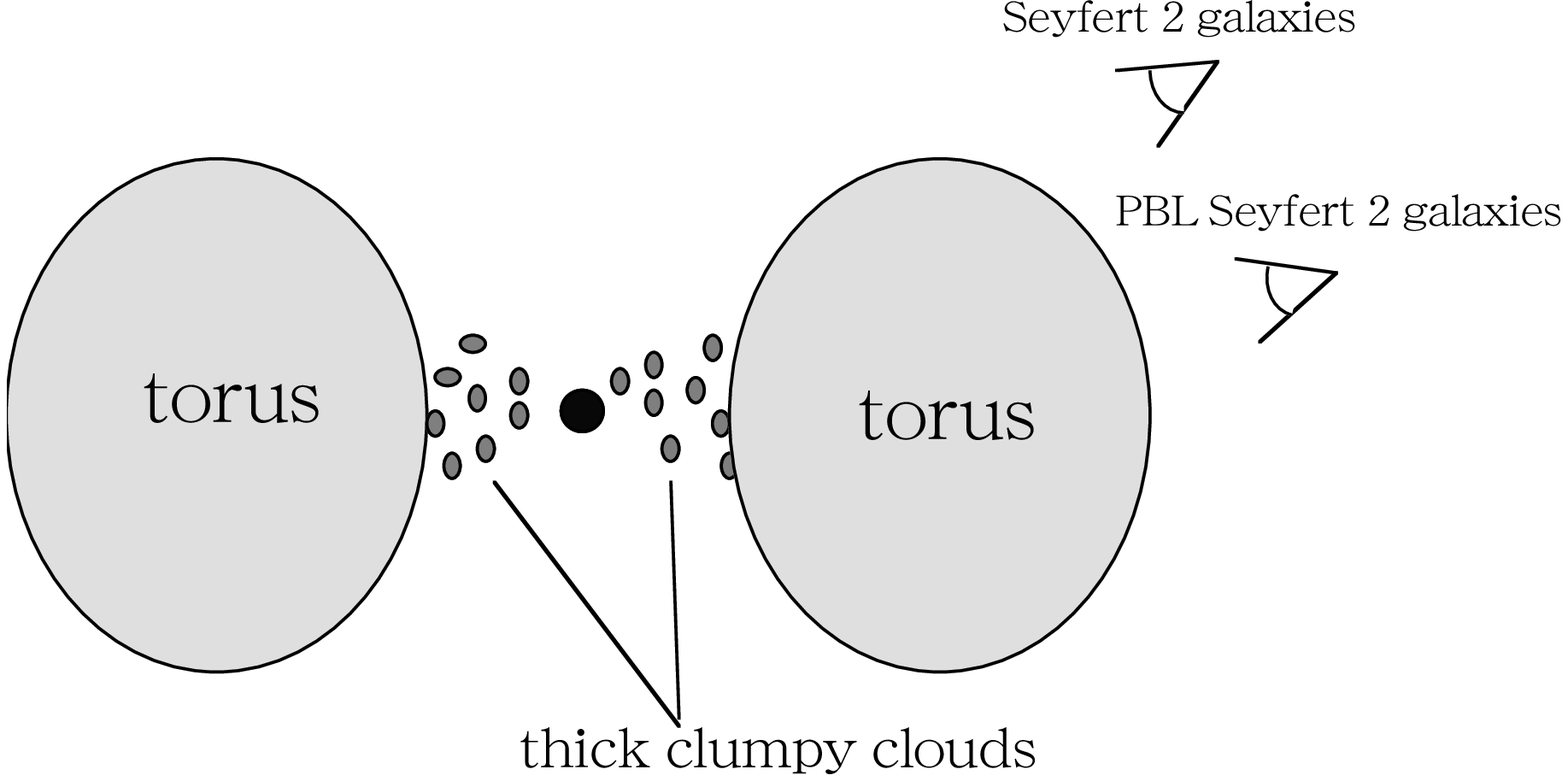}
\caption[figure9a.eps,figure9b.eps]{
Schematic view of Seyfert galaxies proposed in the text.
Left and right figures show anistropic radiation model and
dual absorption model, respectively.
}
\label{figure:9}
\end{figure*}

\end{document}